\documentclass[fleqn,usenatbib]{mnras} %修正
\usepackage{bm}
\usepackage{times}
\usepackage[T1]{fontenc}
\usepackage{ae,aecompl}

\usepackage{graphicx}	% Including figure files
\usepackage{amsmath}	% Advanced maths commands
\usepackage{amssymb}	% Extra maths symbols

\title[CR MHD equations]{Approximate Riemann Solvers for the Cosmic Ray Magnetohydrodynamical Equations}

\author[Y. Kudoh \& T. Hanawa]{
Yuki Kudoh,$^{1}$\thanks{E-mail: kudoyk.aplab@chiba-u.jp (YK)} and
Tomoyuki Hanawa$^{2}$
\\
$^{1}$Department of Physics, Graduate School of Science, Chiba University, 1-33 Yayoi-cho, Inage-ku, Chiba 263-8522, Japan\\
$^{2}$Center for Frontier Science, Chiba University, 1-33 Yayoi-cho, Inage-ku, Chiba 263-8522, Japan}

\date{Accepted XXX. Received YYY; in original form ZZZ}

\pubyear{2016}

\begin{document}
\label{firstpage}
\pagerange{\pageref{firstpage}--\pageref{lastpage}}
\maketitle

\begin{abstract}  
We analyze the cosmic-ray magnetohydrodynamic (CR MHD) equations 
to improve the numerical simulations.  
We propose to solve them in the 
fully conservation form, which is equivalent to the conventional 
CR MHD equations.  In the fully conservation form, the CR energy 
equation is replaced with the CR \lq\lq number" conservation,  where the 
CR number density is defined as the three fourths power of the 
CR energy density. 
The former contains an extra source term, while latter does not.
An approximate Riemann solver is derived from the CR MHD equations
in the fully conservation form.  Based on the analysis, 
we propose a numerical  scheme of which solutions satisfy 
the Rankine-Hugoniot relation at any shock.  
We demonstrate that it reproduces the
Riemann solution derived by \cite{2006MNRAS.367..113P} for a 1D CR hydrodynamic 
shock tube problem.   We compare the solution with those obtained by solving
the CR energy equation.   The latter solutions deviate from the Riemann 
solution seriously, when the CR pressure dominates over the gas pressure 
in the post-shocked gas.
The former solutions converge to the Riemann solution and are of the
second order accuracy in space and time. 
Our numerical examples include an expansion of high pressure sphere
in an magnetized medium.  Fast and slow shocks are sharply resolved
in the example.    We also discuss possible extension of 
the CR MHD equations to evaluate the average CR energy.   
\end{abstract}


%%%%%%%%%%%%%%%%%%%%%%%%%%%%%%%%%%%%%%%%%%%%%%%%%%

%%%%%%%%%%%%%%%%% BODY OF PAPER %%%%%%%%%%%%%%%%%%

\section{Introduction}
Cosmic rays, referred to CRs in the following, are one of the important constituents 
in the interstellar and intergalactic media (see e.g. \citealt{tielens2010physics}).    
The energy density of the CRs is roughly comparable with those of the thermal gas 
and magnetic field.    Thus CRs are expected to 
play a role in the dynamics and magnetic field structures (see e.g. \citealt{longair1994high}).     
Parker (\citeyear{1966ApJ...145..811P, 1967ApJ...149..535P}) pointed out that
CRs enhance the escape of magnetic flux tubes from the Galactic disc through
the Parker instability.   They affect also the termination shock of the solar wind  
(see e.g. \citealt{2013SSRv..176..147F}; \citealt{2013LRSP...10....3P}), 
the galactic winds (see e.g. \citealt{2014A&ARv..22...77B}), 
and cosmological structure formation through shocks (see e.g. \citealt{2006MNRAS.367..113P}; \citealt{2012MNRAS.421.3375V}).

The dynamics of CRs are often taken into account by the CR magnetohydrodynamical
 (MHD) equations.   This fluid approximation is justified because CRs are confined
 by the magnetic field and well scattered by small scale magnetic fluctuations
 (see e.g. \citealt{2013PhPl...20e5501Z}).   CRs are assumed to be an ultrarelativistic gas and hence
 the pressure is assumed to be one third of the energy density in the CR MHD equations.
The CR MHD or CR hydrodynamical (HD) equations have been used
for numerical simulations of the interstellar and galactic scales (see e.g. \citealt{2000A&A...357..268F}; \citealt{2004ApJ...607..828K}; \citealt{2008ApJ...685..105R}; \citealt{2012ApJ...761..185Y}; \citealt{2013ApJ...777L..38H}; \citealt{2014MNRAS.437.3312S}; \citealt{2015A&A...580A.119V}).

The CR MHD equations are similar to the ordinary MHD equations 
and expressed in the conservation form except for the energy equation for CRs.
The energy of CRs is enhanced by compression and consumed by work to the
thermal gas.   Thus the CR energy equation contains a source term proportional 
to spatial derivative of either pressure or velocity.   This source term has been
evaluated separately from the flux in the numerical simulations hitherto.
However, this treatment does not guarantee that the numerical solutions satisfy
the Rankine-Hugoniot relation for a shock.  Although the Rankine-Hugniot relation 
and Riemann solutions have been obtained by \cite{2006MNRAS.367..113P} for the 
CR HD equations, the approximate Riemann solvers have not
been given in the literature.   In other words, the effects of CRs on shocks might not 
be taken into the simulations properly.   Remember that modern HD and MHD
simulations rely on the appropriate approximate Riemann solvers to reproduce
strong shock waves (see e.g. \citealt{toro2009riemann}).  Good approximate Riemann solvers 
capture strong shocks sharply without artificial oscillations of numerical origin.

In this paper, we rewrite the CR MHD equations into the fully conservation form. 
One of them describes the conservation of CR {\it particle number}.  
Here the CR number density is defined to be the 
$ 1/\gamma _\text{cr} $-th power of $ P _\text{cr} $, where $ P _\text{cr} $
and $ \gamma _\text{cr} $ denote the pressure and specific heat ratio of CRs,
respectively.   The Rankine-Hugoniot relation
is easily reconstructed from the fully conservation form. 
We derive the wave properties of the CR MHD equations such as 
the characteristics, the corresponding  right eigenvectors and wave amplitudes. 
They are used to construct the Roe-type approximate Riemann solver
(see e.g. \citealt{toro2009riemann} for the classification of the 
approximate Riemann solvers).
We show that the approximate Riemann solver works well for 1D shock
problems and 2D expansion. The former is used for comparison with 
the exact solution and those obtained with conventional scheme.     
When the CR pressure
is dominant in the post-shocked gas, the solutions obtained with the
conventional schemes do not satisfy the Rankine-Hugoniot relation.
On the other hand, the Roe-type approximate Riemann solver reproduces
the pressure balance mode, a kind of contact discontinuity appearing
in the CR MHD equation, as well as the shock.

This paper is organized as follows.   We derive the fully conservation form 
of the CR MHD equations in Section 2.1.  The Rankine-Hugoniot relation
is derived in Section 2.2.  The approximate Riemann solution is
given in Section 2.3.   Numerical tests are shown in Section 3. 
Section 3.1 is devoted to the 1D CR HD shock tube problem.
Section 3.2 is devoted to the linear wave test,  
while Section 3.3 to the advection of the pressure balance mode.   
Section 3.4 is devoted to the 1D CR MHD shock tube problem,
while Section 3.5 is to 2D CR MHD problem. 
We discuss the shock tube problems 
solved by \cite{2006MNRAS.367..113P} and \cite{2016A&A...585A.138D} 
in Section 4.   We also discuss possible extension of the CR MHD equations
to evaluate the average CR particle energy before conclusion.
This paper is based on our earlier work, \cite{2016JPhCS.719a2021K}, but improved
to achieve the second order accuracy.

%%%%%%%%%%%%%%%%%%%%%%%%%%%%%%%%%%%%%%%%%%%
\section{System of CR MHD equation}

First we introduce the CR MHD equations according to \cite{1990acr..book.....B}
in which the fluid approximation is applied to CRs.
The equation of continuity,
\begin{equation}
\frac{\partial \rho}{\partial t} + \bm{\nabla} \cdot \left( \rho \bm{v} \right)  =0,  \label{mass}
\end{equation}
and the induction equation,
\begin{equation} 
\displaystyle \frac{\partial  \bm{B}} {\partial t} - \bm{\nabla}  \times \left(\bm{v}  \times \bm{B} \right) = 0,  
\label{induction}
\end{equation}
are the same as those of the ordinary MHD equations. 
Here the symbols, $ \rho $, $ \bm{v} $, and $ \bm{B} $ denote the
density, velocity, and magnetic field, respectively.
The CR pressure,
$ P _{\text{cr}} $, is taken into account in the equation of motion, 
\begin{equation}
\displaystyle \frac{\partial}{\partial t} \left( \rho \bm{v} \right) 
+ \bm{\nabla} \cdot \left[ \rho  \bm{vv} + \left( {P}_{\text{g}}  + P_{\text{cr}}
+ \frac{|\bm{B}|^2}{2} \right) \bm{I} - \bm{BB} \right]
= 0 ,  \label{momentum}
\end{equation}
where $ P _{\text{g}} $ and $ \bm{I} $ denote the gas pressure and
the unit tensor, respectively.  Accordingly the equation of energy 
conservation is altered into
\begin{equation}
  \displaystyle \frac{\partial}{\partial t} \left(E \right) 
  + \bm{\nabla} \cdot \left[ \left( E + P _{\text{g}} \right) \bm{v} -  
  \left(\bm{v}  \times \bm{B} \right) \times \bm{B}  \right]  =  - \bm{ v \cdot \nabla} P_{\text{cr} },  
 \label{gasenergy} 
 \end{equation}
 \begin{equation}
 E =  \frac{\rho}{2} |\bm{v}| ^2 + \frac{P _{\text{g}}}{\gamma _{\text{g}} -1}
 + \frac{|\bm{B}| ^2}{2} ,
\end{equation} 
where $ \gamma _{\text{g}} $, denotes the specific heat ratio of the gas. 
The gas is assumed to be an ideal gas with $\gamma_{\text{g}} = 5/3 $ 
throughout this paper except when otherwise noted.

CRs are approximated to be an ideal gas having the constant specific heat ratio, 
$ \gamma _{\text{cr}} = 4/3 $. 
Then the CR energy density, $ E _{\text{cr}} $, is evaluated to be
\begin{equation}
E_{\text{cr}}= \frac{P_{\text{cr}}}{\gamma_{\text{cr}} -1 } .
\end{equation}
The CR energy equation is expressed as
\begin{equation}
\frac{\partial}{\partial t} \left( E_{\text{cr}} \right)
+ \bm{\nabla} \cdot \left[ \left(  E _{\text{cr}} + P _{\text{cr}}  \right) \bm{v} \right]  
=  \bm{v} \cdot \bm{\nabla} P _{\text{cr}} 
- \bm{\nabla} \cdot \bm{F} _{\text{diff}} ,  \label{crenergy}
\end{equation}
\begin{equation}
\bm{F} _{\text{diff}} = - \kappa _\perp \bm{\nabla} E _{\text{cr}}
- \left( \kappa _\parallel - \kappa _\perp \right)
\frac{ \left( \bm{B} \cdot \bm{\nabla} E _{\text{cr}} \right) \bm{B} }{| \bm{B} | ^2}  , 
\end{equation}
where advection, the work to accelerate the gas, and diffusion are
taken into account.   The symbols, $ \kappa _\parallel $
and $ \kappa _\perp $, denote the diffusion coefficients in 
the directions parallel and perpendicular to the magnetic field, respectively.

In the following we analyse the case of no diffusion 
($ \kappa _\parallel = \kappa _\perp = 0 $).   As shown later, the CR MHD
equations are hyperbolic and the characteristic speeds are independent
of the wavelength in this case. Remember that the diffusion is often taken account separately
in numerical simulations by means of the operator splitting.   In other words, 
the diffusion is not taken into account in the construction of approximate 
Riemann solutions.   Thus our approach to the CR MHD equation is orthodox.

%-----2.1------------------------------------------------%
\subsection{Fully conservation Form}

Next we rewrite the CR MHD equations in the fully conservation form.
While equations (\ref{mass})-(\ref{momentum}) are written
in the conservation form, equations (\ref{gasenergy}) and (\ref{crenergy}) are not.
The sum of equations (\ref{gasenergy}) and (\ref{crenergy}) gives us 
the equation of the total energy conservation,
\begin{equation}
\begin{split}
\frac{\partial}{\partial t} \left( E + E _{\text{cr}} \right) 
+ \bm{\nabla} \cdot & \big[ \left( E +  E _{\text{cr}} + P _{\text{g}} + P _{\text{cr}} \right)
\bm{v} \\ 
&~~~~~~~ - \left( \bm{v} \times \bm{B} \right) 
\times \bm{B}  \big] = 0 .
\end{split}
\end{equation}
Thus the CR MHD equations are expressed in the fully conservation form if
equation (\ref{crenergy}) is converted into the conservation form.

For later convenience we introduce the CR number density defined as
\begin{equation}
\rho _{\text{cr}} \equiv P _{\text{cr}} ^{1/\gamma _{\text{cr}}}  ,\label{creos}
\end{equation}
which is equivalent to equation (3.6) in \cite{2006MNRAS.367..113P}.
Equation (\ref{creos}) implies that CRs are approximated by a polytrope gas.
Then equation (\ref{crenergy}) is rewritten in the conservation form,
\begin{equation}
\frac{\partial}{\partial t} \rho _{\text{cr}} + \bm{\nabla} \cdot \left( \rho _{\text{cr}}
\bm{v} \right) = 0 , \label{crnumber}
\end{equation}
when $ \kappa _\parallel = \kappa _\perp = 0 $.  

Note the similarity between equations (\ref{mass}) and (\ref{crnumber}).
From these equations we can derive 
\begin{equation}
\frac{d}{dt} \left( \frac{\rho _{\text{cr}}}{\rho} \right) = 0 . \label{charactoristicCR}
\end{equation} 
For later convenience we introduce the concentration of CRs defined as
\begin{equation}
\chi = \frac{\rho _{\text{cr}}}{\rho} .
\end{equation}

When all the variables depend only on $ t $ and $ x $ (1D), the
CR MHD equations are expressed in the vector form,
\begin{equation}
\displaystyle \frac{\partial \bm{U}}{\partial t}+ \frac{\partial \bm{F}}{\partial x} =0, 
\label{vector1}
\end{equation}
\begin{equation}
\bm{U}=                             \begin{bmatrix} 
\rho \\  \rho v_x \\ \rho v_y \\  \rho v_z \\  B_x \\ B_y \\ B_z \\ \rho H - P_{\text{T}} + 
\displaystyle \frac{|\bm{B}|^2}{2} \\   \rho  \chi 
                                         \end{bmatrix} , \label{vector2} 
\end{equation}
\begin{equation}
\bm{F} =                             \begin{bmatrix} 
\rho v_x \\ \displaystyle \rho v_x^2 + P_{\text{T}} + \frac{|\bm{B}|^2 }{2} - B_x^2 \\ \rho v_x v_y - {B_x B_y} \\ \rho v_x v_z - B_x B_z \\ 0 \\ v_x B_y - v_y B_x \\ v_x B_z - v_z B_x \\  \rho H v_x + |\bm{B}|^2 v_x - (\bm{v} \cdot \bm{B} ) B_x  \\  \displaystyle \rho \chi v_x
                                        \end{bmatrix}     , \label{vector3}
\end{equation}
\begin{equation}
H = \frac{\bm{v} ^2}{2} + \frac{\gamma _{\text{g}}}{\gamma _{\text{g}} - 1} \frac{P _{\text{g}}}{\rho}
+ \frac{\gamma _{\text{cr}}}{\gamma _{\text{cr}} - 1} \frac{P _{\text{cr}}}{\rho} ,
\end{equation}
\begin{equation}
P _{\text{T}} = P _{\text{g}} + P _{\text{cr}} .
\end{equation}
where $\bm{U}$ and $\bm{F}$ denote the state and flux vectors,
respectively.  We use this vector form to derive the Rankine-Hugniot relation and
Riemann solution.

%-----2.2-----------------------------------------------%
\subsection{Rankine-Hugoniot relation}

In this subsection we derive the Rankine-Hugoniot relation, i.e, the
jump condition at a shock front, using  the CR MHD equations
in the conservation form, equations (\ref{vector1}),
(\ref{vector2}) and (\ref{vector3}).
For simplicity we restrict ourselves to a stationary plane shock.
In other words, we observe a small area around a shock wave in
the comoving frame.  Furthermore, the wave front is assumed
to be normal to the $ x $-direction in the Cartesian coordinates.
Since the temporal change vanishes, then the flux, $ \bm{F} $, should be continuous across the shock 
front, 
\begin{eqnarray}
\left[ \rho v _x \right] & = & 0 , ~~~~ \label{massc} \\
\left[ \rho v _x ^2 + P _{\text{T}} + \frac{B _y ^2 + B _z ^2 - B _x ^2}{2} \right]
& = & 0 ,~~~~ \label{vxc} \\
\left[ \rho v _x v _y - B _x B _y \right] & = & 0, ~~~~\label{vyc} \\
\left[ \rho v _x v _z - B _x B _z \right] & = & 0, ~~~~\label{vzc} \\
\left[ v _x B _y - v _y B _x \right] & = & 0, ~~~~ \label{Byc} \\
\left[ v _x B _z - v _z B _x \right] & = & 0, ~~~~ \label{Bzc} \\
\left[ \rho H v_x + \left( B _y ^2 + B _z ^2 \right) v _x - \left( v _y B _y + v _z B _z \right)
B _x \right] & = & 0, ~~~~ \label{Hc} \\
\left[ \rho \chi  v _x \right] & = & 0, ~~~~ \label{CRc}
\end{eqnarray}
where the symbol $[ \cdot]$ denotes the jump across the discontinuity. 
Here, the subscripts, $ x $, $ y $, and $ z $ denote 
the $ x $-, $ y $-, and $ z $-components, respectively. 
These conditions are the same as those obtained by \cite{2006MNRAS.367..113P} and
almost the same as those for the MHD equations
except for equation (\ref{CRc}): equations (\ref{vxc}) and (\ref{Hc}) are
modified to include $ P _{\text{cr}} $ and $ E _{\text{cr}} $.
Equation (\ref{CRc}) denotes the continuity of  the CR 
number flux.  
Combining equations (\ref{massc}) through (\ref{CRc}) we obtain
\begin{equation}
 \frac{\rho _{\text{cr}}}{\rho}  = \chi =  \text{const.}, \label{CRc2}
\end{equation}
across the shock.  This means that the CR pressure 
changes only through the gas compression or expansion.

%-----2.3----------------------------------------------------------------%
\subsection{Elementary Wave Solutions of the Riemann Problem}

First we examine the CR HD equations,  since inclusion of CRs 
changes the equation of state but does not alter the induction equation.
Furthermore we consider a 1D flow in which the 
$ y $- and $ z $-components vanish.
Then the state and flux vectors are expressed as
\begin{equation}
\bm{U}=                             \begin{bmatrix} 
\rho \\  \rho v _x \\ \rho H - P_{\text{T}} \\
\rho  \chi 
                                         \end{bmatrix} ,  \label{CRHD1DU}
\end{equation}
\begin{equation}
\bm{F}=                             \begin{bmatrix} 
\rho v _x \\  \rho v _x ^2 + P _{\text{T}} \\ \rho H v_x \\
\rho  \chi v_x
                                         \end{bmatrix} . \label{CRHD1DF}
\end{equation}
As shown later, inclusion of the tangential velocity and magnetic
field are straight forward.

CRs increase the total pressure and hence the sound speed.
The total pressure is expressed as
\begin{equation}
P _{\text{T}}  = P _{\rm g} + \left( \chi \rho \right) ^{\gamma _{\text{cr}}} ,
\end{equation}
as shown in the previous subsection.
Then the adiabatic sound speed is evaluated to be
\begin{equation}
a  = 
\left[ \left(\frac{\partial P_{\text{g}}}{\partial \rho}\right)_{s} + 
\left(\frac{\partial P_{\text{cr}}}{\partial \rho}\right)_{\chi} \right]^{1/2} = \left( \frac{\gamma_{\text{g}} P _{\text{g}} + \gamma _{\text{cr}} P _{\text{cr}}}{\rho} 
 \right) ^{1/2},
\end{equation}
\begin{equation}
s = \ln P _{\text{g}} - \gamma _{\text{g}} \ln \rho ,
\end{equation}
where $ s $ denotes the entropy, since $d \chi/dt =0 $ (equation \ref{charactoristicCR}).  Note that
the entropy is constant
\begin{equation}
\frac{ds}{dt} = 0 ,
\end{equation}
for a given gas element except for increase at a shock.

The characteristics of the CR HD equations are the 
eigenvalues of the Jacobian matrix,
\begin{equation}
\bm{A} \equiv \partial \bm{F}/\partial \bm{U} .
\end{equation}
We obtain four characteristics,
\begin{equation}
\lambda _{1,4} = v _x \pm a ,
\end{equation}
\begin{equation}
\lambda _{2,3} = v _x ,
\end{equation}
after some algebra.
The former denotes the sound wave while the latter does
advection.   The additional advection mode corresponds to
conservation of the CR concentration 
(equation \ref{CRc2}) and hence to the pressure balance mode.  
The others are the same as those
in the ordinary HD equations except for the change in
the sound speed.

The spatial derivative of $ \bm{U} $ and $ \bm{F} $ are
decomposed into waves.  They are expressed as
\begin{equation}
\frac{\partial \bm{U}}{\partial x} = \sum _{i=1} ^4 
w _i \bm{r} _i , \label{dcomps}
\end{equation} 
\begin{equation}
\frac{\partial \bm{F}}{\partial x} = \sum _{i=1} ^4 \lambda _i  
w _i \bm{r} _i ,
\end{equation} 
where
\begin{equation}
w_{1,4}=\displaystyle \frac{1}{2 a^2} \left( \frac{\partial P_{\text{T}}}{\partial x} \pm {\rho}  a \frac{\partial v_x}{\partial x}  \right),  \label{w14}
\end{equation}
\begin{equation}
w_2=\displaystyle \frac{\partial \rho}{\partial x} - \frac{1}{ a^2 } \frac{\partial P_{\text{T}} }{\partial x} = - \frac{P_{\text{g}}}{a^2} \frac{\partial s}{\partial x} - \frac{\rho \chi^{-1}}{a^2} \left( \frac{\partial P_{\text{cr}}}{\partial \rho} \right)_{\chi} \frac{\partial \chi}{\partial x} ,~~
\end{equation}
\begin{equation}
w_3=\frac{\partial \rho_{\text{cr}}}{\partial x} - \chi \frac{\partial \rho}{\partial x}= \rho \frac{\partial \chi }{\partial x} ,
%w_4=\displaystyle  \frac{1}{2 \bar{c}_s^2}(\Delta P_{\text{T}} - \bar{\rho} \bar{c}_s \Delta v_x).
\end{equation}
 \begin{equation}
\displaystyle 
\bm{r_{1,4}}=                             \begin{bmatrix} 
\displaystyle 1 \\ {v_x} \pm {a} \\  {H} \pm a {v}_x \\   \chi 
                                         \end{bmatrix}  
,~~\bm{r_2}=                             \begin{bmatrix} 
\displaystyle 1 \\ {v}_x  \\ \displaystyle \frac{{v}_x^2}{2} + \chi \zeta \\   \chi
                                         \end{bmatrix}  
,~~\bm{r_3}=                             \begin{bmatrix} 
\displaystyle 0 \\ 0  \\ \zeta \\  1 
                                         \end{bmatrix} ,  \label{dcompe}
%,~~\bm{r_4}=                             \begin{bmatrix} 
%\displaystyle 1 \\ {v_x} - a \\  {H} -  a v_x \\ 
%\chi                                       \end{bmatrix},
\end{equation} 
\begin{equation}
\zeta = \frac{ \gamma_{\text{g}} - \gamma_{\text{cr}} }{(\gamma_{\text{g}} -1)(\gamma_{\text{cr}} - 1)} \frac{d P_{\text{cr}}}{d \rho_{\text{cr}}}.
\end{equation}
Here the symbol, $ w _i $, denotes the amplitude of the $i$-th wave.
Inclusion of CRs introduces a new Riemann invariant, $ \chi $, and
change the other Riemann invariants only slightly.  The symbol,
$ \zeta $, denotes the change in the energy density due to the
pressure balance mode (\citealt{1995ApJ...442..822W}).   The energy density depends on the
proportion of $ P _{\text{cr}} $  for a given $ P _{\text{T}}$.

Equations (\ref{dcomps}) through 
(\ref{dcompe})  are used to construct the \cite{Roe.1981.ARS} type 
approximate Riemann solvers for numerical simulations.  
Further details are given in Appendix A.

Also the exact Riemann solutions are derived from equations
(\ref{dcomps}) through (\ref{dcompe}).   Here the Riemann solution
means that the solution of equation (\ref{vector1}), when
the initial condition  is expressed as
\begin{equation}
\bm{U} = 
\begin{cases}
\bm{U} _{\text{L}} & x < 0 \\
\bm{U} _{\text{R}} & x \ge 0
\end{cases} ,
\end{equation}
at $ t  = 0 $.  The Riemann solution is obtained from the conditions,
\begin{equation}
w _i = 0 ,
\end{equation}
except at discontinuities.  However, the algorithm to obtain the
complete solution is lengthy even for the ordinary HD equations, as shown in \cite{toro2009riemann}.  
\cite{2006MNRAS.367..113P} showed
the Riemann solution for the case in which a shock propagates 
rightward and a rarefaction wave propagates leftward.

\begin{figure}  %%% Section 2.3 -- Fig. 1 %%%%%%%%%
\centering
\begin{center}
\includegraphics[width=\columnwidth]{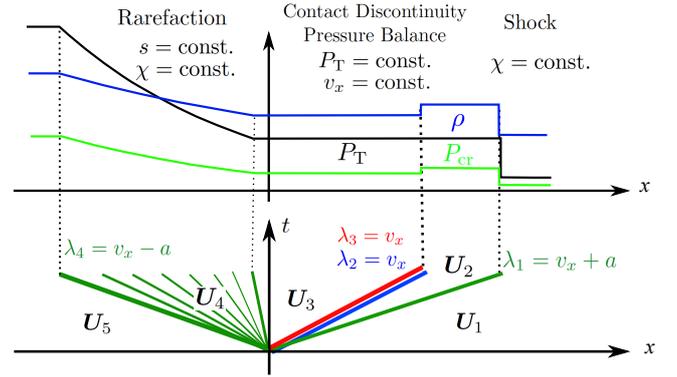}
\end{center}
\caption{\label{shocktube} Illustration of the Riemann solution for 
a CR HD shock tube problem.  The density, total pressure and CR pressure are
shown as a function of $ x $ in the upper half.  The lower half
denotes the characteristics.}
\end{figure}   %%%%%%%%%%%%%%%%%%%%%%%%%%

Fig.~\ref{shocktube} illustrates an example of the Riemann solution.
In this solution a shock wave travels at the speed, $ \lambda _1 $,
while the contact discontinuity and the pressure balance
mode at $ \lambda _2 (= \lambda _3)$.   The rarefaction wave
has the head and tail.    The state vector is uniform in each region 
separated by the
characteristics except between the head and tail of the rarefaction
wave.  We refer to regions 1 through 5 from right to left in the
figure.  The Rankine-Hugoniot relation is applied to the jump 
between regions 1 and 2.   The total pressure and velocity
are continuous at the boundary between regions 2 and 3, while
the entropy and CR concentration are not.   The velocity gradient
in region 4 is obtained from
\begin{equation}
\frac{dv _x}{dx} = \frac{1}{\rho a} \frac{dP _{\rm T}}{dx} ,
\end{equation}
$ s = \text{const.} $, $ \chi = \text{const.} $, and $ \lambda _4 = v_x - a $.
The state vector remains at the initial value,
$ \bm{U} _1 = \bm{U} _{\text{R}} $ and 
$ \bm{U} _5 = \bm{U} _{\text{L}} $, in regions 1 and 5.
In short the Riemann solution changes only quantitatively by
inclusion of CRs.

Next we consider the Riemann solution of the CR MHD equations.
The ordinary MHD equations have 7 characteristics: three pairs of
the fast, slow, and Alfv\'en waves in addition to the entropy wave.  
CRs add the pressure balance mode and modify the phase speeds  
of the fast and slow waves through the change in the sound speed.
After some algebra, we have found the following formulae 
for $ \lambda _i $,  $ w _i $, and $ \bm{r}_{i} $ for the CR MHD equations 
using the formulation given by \cite{1995ApJ...442..228R} for the
ordinary  MHD equations.
% eigen value
\begin{equation}
\lambda_{1,8}=v_x \pm c_f,
\end{equation}
\begin{equation}
\lambda_{3,6}=v_x \pm c_s,
\end{equation}
\begin{equation}
\lambda_{4,5}=v_x, ~~ \lambda_{2,7}= v_x \pm v_{\text{A}x}, 
\end{equation}
%  amplitude Fast
\begin{equation}
\begin{split}
w_1 &+ w_8  = 
\frac{\alpha_f}{c_f^2} \left( \frac{\partial P_{\text{T}}}{\partial x} +  B_y \frac{\partial B_y}{\partial x} +B_z \frac{\partial B_z}{\partial x} \right) \\
       &+  \Bigg[ \frac{\alpha_s \sqrt{\rho}}{a^2 c_f} \bigg\{ \left( \gamma_{\text{g}} -1 \right) c_s^2 - \left( \gamma_{\text{g}} -2 \right) a^2  \bigg\}  \\
       &~~~~ + \frac{\alpha_f}{c_f^2} \left( \gamma_{\text{g}} -2 \right) \sqrt{B_y^2+B_z^2}  \Bigg] \left( \beta_y \frac{\partial B_y}{ \partial x} + \beta_z \frac{\partial B_z}{\partial x } \right),
\end{split}
\end{equation}
\begin{equation}
w_1-w_8 = \frac{\alpha_f}{c_f} \rho \frac{\partial v_x}{\partial x} - \frac{\alpha_s c_s}{c_f a} \rho ~ \text{sgn} \left( B_x\right)  \left( \beta_y \frac{\partial B_y}{\partial x } + \beta_z \frac{\partial B_z}{\partial x}\right),
\end{equation}
% amplitude Slow
\begin{equation}
\begin{split}
w_3 &+ w_6  = 
\frac{\alpha_s}{a^2} \left( \frac{\partial P_{\text{T}}}{\partial x} +  B_y \frac{\partial B_y}{\partial x} +B_z \frac{\partial B_z}{\partial x} \right) \\
       &+  \Bigg[ \frac{\alpha_f \sqrt{\rho}}{a^2 c_f} \bigg\{ \left( \gamma_{\text{g}} -2 \right) a^2 - \left( \gamma_{\text{g}} -1 \right) c_f^2  \bigg\}  \\
       &~~~~ + \frac{\alpha_s}{a^2} \left( \gamma_{\text{g}} -2 \right)\sqrt{B_y^2+B_z^2}   \Bigg] \left( \beta_y \frac{\partial B_y}{ \partial x} + \beta_z \frac{\partial B_z}{\partial x } \right),
\end{split}
\end{equation}
\begin{equation}
w_3-w_6 = \frac{\alpha_s v_{\text{A}x} }{c_f a} \rho \frac{\partial v_x}{\partial x} + \frac{\alpha_f }{a} \rho ~ \text{sgn}  \left( B_x\right)  \left( \beta_y \frac{\partial B_y}{\partial x } + \beta_z \frac{\partial B_z}{\partial x}\right),
\end{equation}
% amplitude Entropy
\begin{equation}
w_4 = \frac{\partial \rho}{\partial x} - \alpha_f (w_1+ w_8) - \alpha_s (w_3+ w_6),
\end{equation}
% amplitude Pressure Balance
\begin{equation}
w_{5}= \rho \frac{\partial \chi}{ \partial x},
\end{equation}
% amplitude Alfven
\begin{equation}
\begin{split}
w_{2,7} &=\frac{1}{2} \Bigg[ \mp \rho \left( \beta_z \frac{\partial v_y}{\partial x} - \beta_y \frac{\partial v_z}{\partial x} \right) \text{sgn} \left( B_x \right) \\
           &~~~~~~~+\sqrt{\rho} \left( \beta_z \frac{\partial B_y}{\partial x} - \beta_y \frac{\partial B_z}{\partial x} \right)  \Bigg],
\end{split}
\end{equation}

% right eigenvalue  Entropy, PressureBarance, Alfven
% Fast
\begin{equation}
\bm{r}_{1,8}=   \begin{bmatrix} 
\displaystyle \alpha_f \\ \alpha_f (v_x \pm c_{f})  \\ \alpha_f v_y \mp \alpha_s \beta_y v_{Ax} \text{sgn}({B_x}) \\ \alpha_f v_z \mp \alpha_s \beta_z v_{Ax} \text{sgn}({B_x}) \\ \displaystyle \frac{\alpha_s \beta_y c_{f} }{\sqrt{\rho}} \\ \displaystyle \frac{ \alpha_s \beta_z c_{f} }{\sqrt{\rho}} \\ R_{1,8} \\ \displaystyle \alpha_f  \chi 
                        \end{bmatrix},
\end{equation}

% Slow
\begin{equation}
\bm{r}_{3,6}=   \begin{bmatrix} 
\displaystyle  \alpha_s   \\   \alpha_s (v_x \pm c_{s})    \\    \alpha_s v_y \pm \alpha_f \beta_y a~ \text{sgn}({B_x}) \\ \alpha_s v_z \mp \alpha_f \beta_z a~ \text{sgn}({B_x}) \\ \displaystyle - \frac{\alpha_f \beta_y a^2}{c_{f} \sqrt{\rho} } \\ \displaystyle  - \frac{\alpha_f \beta_z a^2}{c_{f} \sqrt{\rho} } \\ \displaystyle R_{3,6} \\  \displaystyle \alpha_s  \chi 
                        \end{bmatrix},
\end{equation}
\begin{equation}
\bm{r}_{2,7}=   \begin{bmatrix} 
\displaystyle 0 \\ 0  \\ \mp \beta_z  \text{sgn}({B_x}) \\ \pm \beta_y \text{sgn}({B_x}) \\ \beta_z  / \sqrt{\rho} \\ - \beta_y / \sqrt{\rho} \\ R_{2,7} \\ 0 
                        \end{bmatrix}, ~~
\bm{r}_{4}=   \begin{bmatrix} 
\displaystyle 1 \\ v_x \\ v_y \\ v_z \\ 0 \\ 0 \\ R_{4} \\ \chi 
                        \end{bmatrix},~~
\bm{r}_{5}=   \begin{bmatrix} 
\displaystyle 0 \\ 0 \\ 0 \\ 0 \\ 0 \\ 0 \\ \zeta \\ 1 
                        \end{bmatrix},
\end{equation}
where,
\begin{equation}
\displaystyle R_{4}=\frac{|\bm{v}|^2}{2} + \chi \zeta , ~~~
R_{2,7}= \mp \left( \beta_z v_y - \beta_y v_z \right) \text{sgn} (B_x) ,
\end{equation}
\begin{equation}
\begin{split}
R_{1,8} &=\alpha_f \left\{ R_4 \pm c_f v_x + \frac{c_f^2}{\gamma_{\text{g}}-1 } + \frac{ \gamma_{\text{g}}-2 }{ \gamma_{\text{g}}-1 } \left( c_f^2 - {a}^2 \right) \right\} \\
           &~~\mp \alpha_s v_{Ax} ~ \text{sgn} (B_x) \left( \beta_y v_y + \beta_z v_z \right), 
\end{split}
\end{equation}
\begin{equation}
\begin{split}
R_{3,6} & =\alpha_s \left\{ R_4 \pm c_s v_x + \frac{c_s^2}{\gamma_{\text{g}}-1 } + \frac{ \gamma_{\text{g}}-2 }{ \gamma_{\text{g}}-1 } \left( c_s^2 - {a}^2 \right) \right\} \\
          &~~ \pm \alpha_f {a} ~ \text{sgn} (B_x) \left( \beta_y v_y + \beta_z v_z \right),
\end{split}
\end{equation}
\begin{equation}
v_{Ax}^2 = \frac{B_x^2}{\rho}, ~~~ a_{\ast}^2 = a^2 + \frac{B_x^2+ B_y^2 +B_z^2}{\rho}  ,
\end{equation}
\begin{equation}
c_{f,s}^2 = \frac{1}{2} \left( a_{\ast}^2 \pm \sqrt{  a_{\ast}^4 - 4a^2 v_{Ax}^2}  \right) ,
\end{equation}
\begin{equation}
\displaystyle \alpha_f=\frac{\sqrt{c_f^2-v_{Ax}^2}}{\sqrt{c_f^2-c_s^2}},~~~
\alpha_s=\frac{\sqrt{c_f^2-a^2}}{\sqrt{c_f^2-c_s^2}},~~~
\alpha_f^2+\frac{v_{Ax}^2}{c_f^2} \alpha_s^2 =1,
\end{equation}
\begin{equation}
\displaystyle \beta_y=\frac{B_y}{\sqrt{B_y^2+B_z^2}},~~~
\beta_z=\frac{B_z}{\sqrt{B_y^2+B_z^2}},~~~
\beta_y^2 + \beta_z^2 = 1.
\end{equation}

The above formulae are also used to construct the Roe type
approximate Riemann solutions.  The Roe average is given
in Appendix A.  It is difficult to obtain the exact CR MHD Riemann solution 
although it may not be impossible.

% Section. 3 %%%%%%%%%%%%%%%%%%%%%%%%%%%%%%%%%%%%
\section{Numerical Tests} 

We examine whether we can reproduce the Riemann solution 
given in the previous section in numerical simulations.
CR MHD simulations thus far have solved equation (\ref{crenergy})
instead of equation (\ref{crnumber}).   We show that a solution
of equation (\ref{crenergy}) does not satisfies the Rankine-Hugoniot
relation for a 1D CR HD shock tube problem, while equation 
(\ref{crnumber}) does.   We also demonstrate equation (\ref{crnumber})
provides appropriate solutions for 1D CR MHD shock tube problem and
2D CR MHD expansion problem.
 
We  use the explicit scheme of the second order accuracy in space and time
except when otherwise noted.
The  second order accuracy in space is achieved by the Monotone 
Upstream-Centered Schemes for Conservation Laws (MUSCL) with
characteristics (\citealt{1979JCoPh..32..101V}; and see e.g. \citealt{toro2009riemann}). 
 When we solve equation (\ref{crenergy}), we interpolate
the primitive variables, $ \rho $, $ P _{\text{g}} $, $ P _{\text{cr}} $, 
$ v _x  $,  $ v _y $, $ v _z $, $ B _x $, $ B _y $, and $ B _z $.
When we solve the fully conservative form of the CR MHD equations, 
we interpolate the wave amplitudes, $ w _i $,
instead of the primitive variables.   The interpolated variables are chosen
to achieve the second order accuracy in solving the pressure balance mode.  
The second order accuracy in time is achieved by the two stage Runge-Kutta method.
All the test problems are solved on a uniform cell width and with a constant
time step.

%-- SubSection 3.1 -------------------------------------------------------------------%
\subsection{1D CR HD shock tube problem}

\begin{table}  %% Section 3.1 ---- Table 1 %%%%%%
\caption{\label{ex} The initial state in the CR HD shock tube problem.}
\begin{center}
\begin{tabular}{rllll}
\hline
~~ & ~$\rho$ & $v_x$ & $P_{\text{g}}$ & $P_{\text{cr}}$ \\ \hline
Left $(x < 0)$ & 1.0 & 0.0 &  $ 2.0  $& $1.0 $\\ 
Right $(x\geq 0)$ & 0.2 & 0.0  & $ 0.02 $ & $ 0.1 $ \\
\hline
\end{tabular}
\end{center}
\end{table}%%%%%%%%%%%%%%%%%%%%%%%%

In this subsection, we solve a CR HD shock tube problem with four
different schemes.  Two of them solve equation (\ref{crnumber})
while the others two not.   The initial state is summarized in 
Table \ref{ex}.   The left-hand side has a higher pressure and hence the
shock wave runs rightward with the speed, $ 2.369 $.
The head of the rarefaction wave runs leftward with the speed,
$ - 2.160$.   The spatial resolution is $ \Delta x  = 1/128 $ and the
time step is fixed at $ \Delta t =2.0 \times 10^{-3} $.  Accordingly the Courant-Friedrich-Lewy (CFL) number
is 0.780 in the simulations shown below.

\begin{figure*}   %% Section 3.1 ---- Fig 2 %%%%%%%%%%%%
\begin{center}
\includegraphics[width=0.9\textwidth]{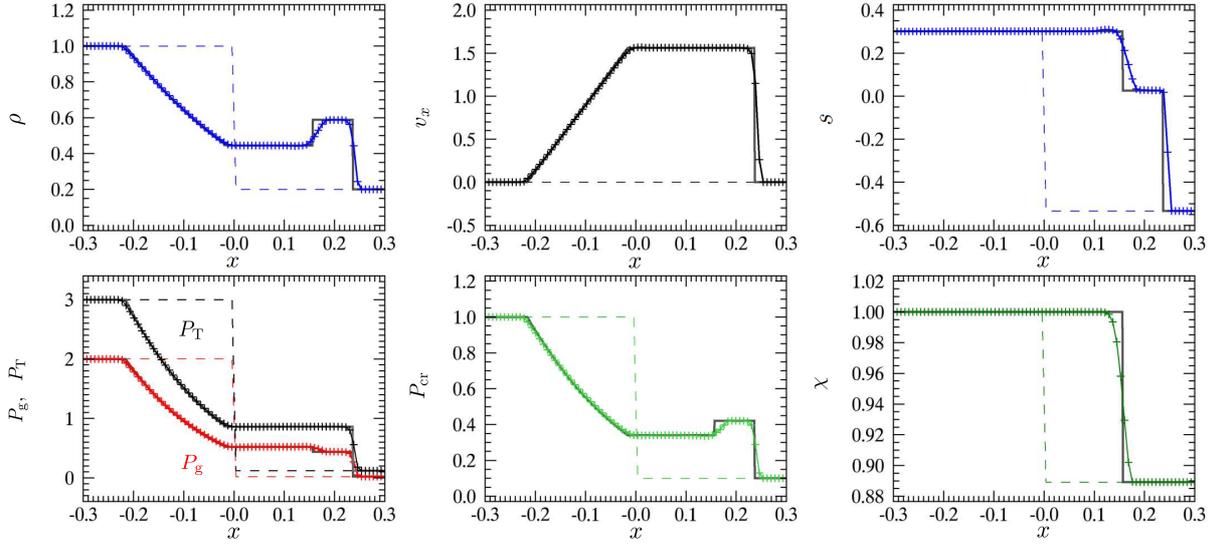}
\end{center}
\caption{\label{CRHDshock} 
The numerical solution of the CR HD shock tube problem compared with the analytic
Riemann solution (thin black solid) and the initial state (dashed).  
The numerical solution (symbols) is 
of the second order accuracy and obtained with  the Roe-type Riemann solver.
 Top: $ \rho $, $ v _x $ and $s$ from left to right.  Bottom: $ P _{\text{g}} $,
$ P _{\text{T}} $, $ P _{\text{cr}} $ and $ \chi $ from left to right.}
\end{figure*}%%%%%%%%%%%%%%%%%%%%%%%%%%%%

First we solve the CR HD equations in the fully conservation form,
in which the state and flux vectors are given by equations
(\ref{CRHD1DU}) and (\ref{CRHD1DF}),  with the Roe-type approximate Riemann solver. 
The solution at $ t = 0.1 $ is shown in Fig.~\ref{CRHDshock}.
The top panels show the density, velocity, and entropy from left
to right, respectively.  The bottom left-hand panel shows $ P _{\text{g}} $ 
by the red curve and $ P _{\text{T}} $ by the black curve.  The bottom
central panel denotes $ P _{\text{cr}} $ by the yellow green while 
the bottom right-hand panel does CR concentration ($ \chi $) by the green.
The plus symbols denote the data points.
The black curves without symbols denote the Riemann solution,  
while the dashed lines do the initial state.

The shock front and rarefaction wave are reproduced well in 
Fig. \ref{CRHDshock}.   The contact discontinuity ($ x = 0.156 $) is also
reproduced well.   Note that CR concentration, $ \chi $, 
is constant across the shock and changes at the contact 
discontinuity.

\begin{figure*} %%%% Subsection 3.1 ---- Fig 3 %%%%%%%%%%%%%%
\begin{center}
\includegraphics[width=0.9\textwidth]{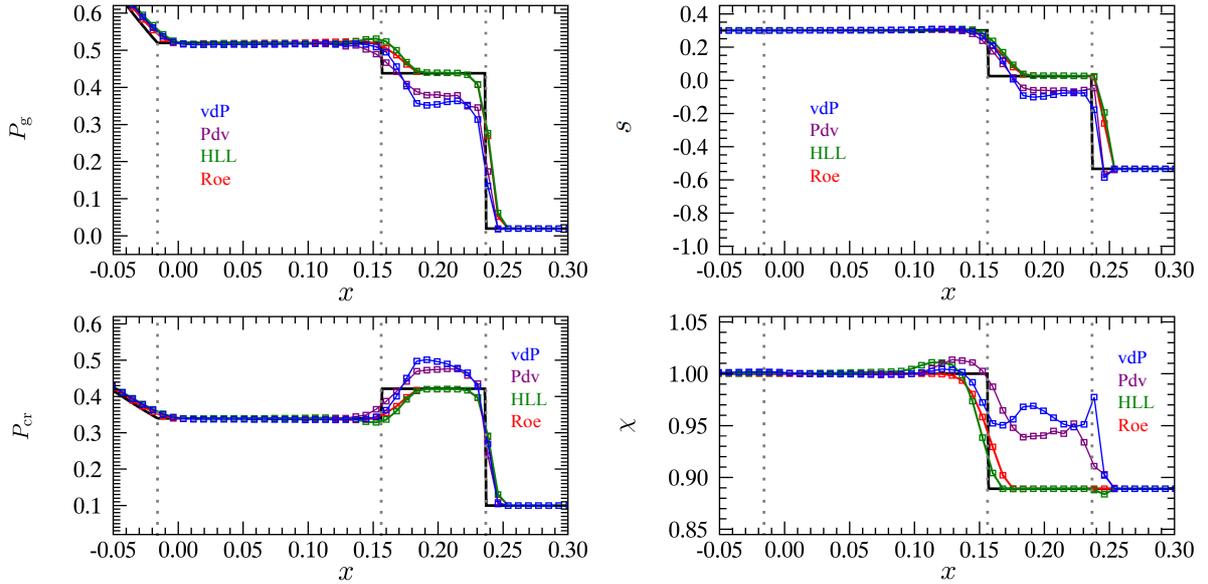}
\end{center}
\caption{\label{CRHD2nd} 
Comparison of the numerical solutions with the 2nd order accuracy in the CR HD shock tube problem.
The green and red curves denote the solutions obtained with the CR HD equations in the
conservation form while the blue and purple curves do those obtained with the equations
with source terms. 
Top: the gas pressure (left) and the entropy (right).  
Bottom: the CR pressure (left) and the CR concentration (right).
See the main text for further details.}
\end{figure*}  %%%%%%%%%%%%%%%%%%%%%%%%%%%%%%%%

\begin{figure*} %%%%  Subsection 3.1 ---- Fig 4 %%%%%%%%%%%% 
\begin{center}
\includegraphics[width=0.9\textwidth]{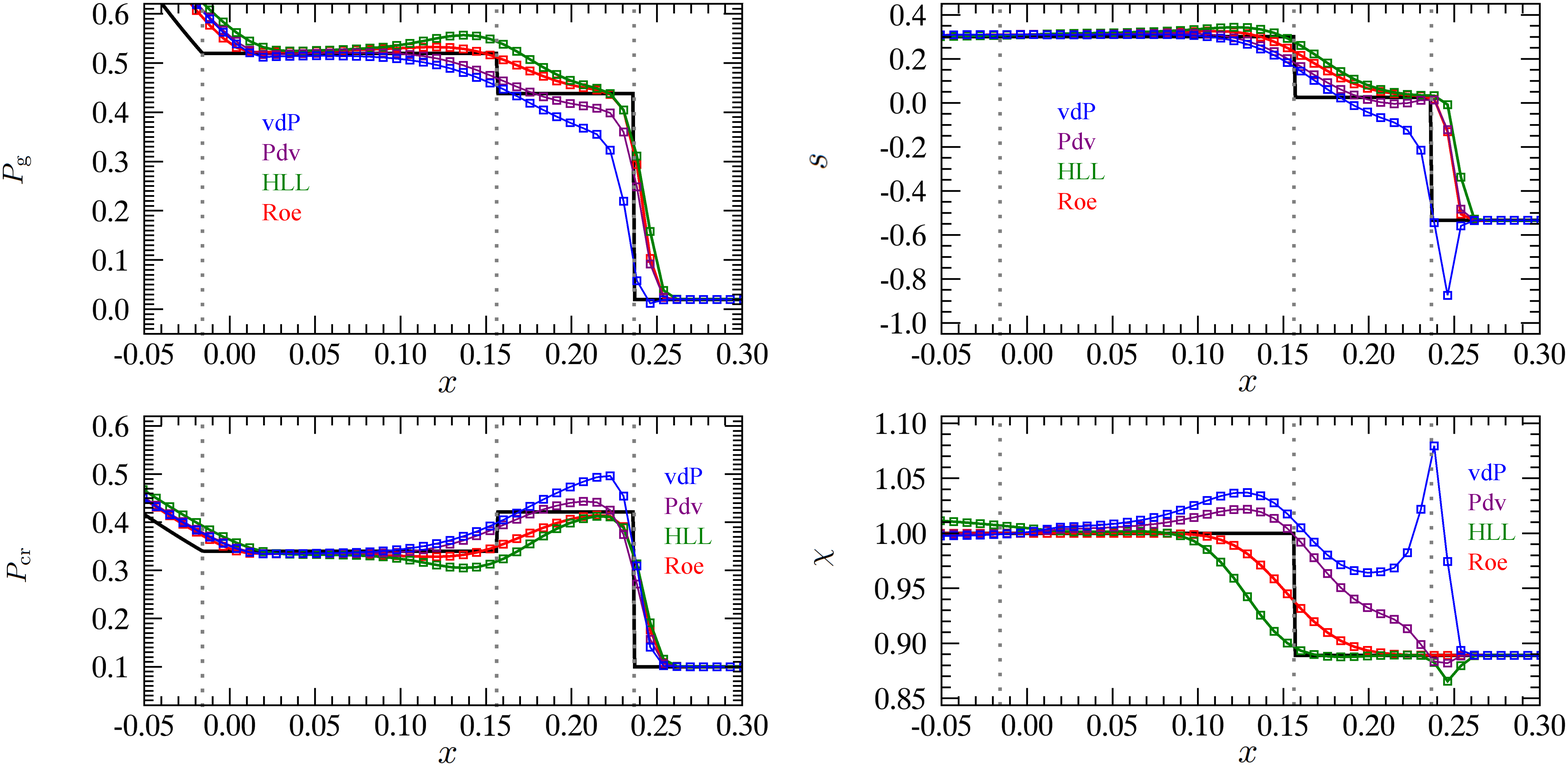}
\end{center}
\caption{\label{CRHD1st} 
The same as Fig. \ref{CRHD2nd} but for the numerical solution of 
the 1st order accuracy.}
\end{figure*}   %%%%%%%%%%%%%%%%%%%%%%%%%%%%%%%

Fig. \ref{CRHD2nd} compares this solution with three other solutions obtained
with different schemes.   Each panel  shows the enlargement around
the interval between the shock front and contact discontinuity where
the differences are large.  The top left and right panels denote
$  P _{\text{g}} $ and $ s $ at $ t = 0.1 $, respectively,  while 
the bottom left and right ones $ P _{\text{cr}} $ and 
$ \chi $, respectively.   The red curves
denote the solution shown in Fig. \ref{CRHDshock}.    

The green curves denote the solution obtained with the Harten-Lax-van Leer (HLL) scheme 
(see e.g. \citealt{toro2009riemann})
applied to the fully conservation form of the CR HD equations. 
The maximum and minimum characteristics are evaluated to be
$ v _x \pm a $ in the HLL scheme.
It provides a good approximation around the shock
while the deviation from the Riemann solution is large around the
contact discontinuity.   

The blue curves denote the solution of equation (\ref{crenergy}).
The source term, the right hand side, is evaluated by the central
difference.  The other equations are solved with the HLL 
scheme.   This scheme is named vdP scheme after the source term.   
Similar schemes are employed in \cite{2004ApJ...607..828K}.
This solution is different from the Riemann solution between the
shock front and contact discontinuity.   The gas pressure is
lower and cosmic ray pressure is higher.

The purple curves denote solution obtained by solving
\begin{equation}
%\frac{\partial}{\partial t} \left( \frac{P _{\text{cr}}}{\gamma _{\text{cr}} - 1}
%\right) + \bm{\nabla} \cdot \left( \frac{P _{\text{cr}}}
%{\gamma _{\text{cr}} - 1} \bm{v} \right) = - P _{\text{cr}} \bm{\nabla} \cdot \bm{v}  ,
\frac{\partial}{\partial t} \left( E_{\text{cr}} \right) + \bm{\nabla} \cdot \left( E_{\text{cr}} \bm{v} \right) 
= - P _{\text{cr}} \bm{\nabla} \cdot \bm{v}  ,
\end{equation}
which is derived from equation (\ref{crenergy}).   The right hand
side is evaluated by the central difference and the other equations are
solved with the HLL scheme, which we call Pdv scheme in the following.
Similar schemes are employed in \cite{2003A&A...412..331H}, \cite{2012ApJ...761..185Y}, 
and \cite{2016A&A...585A.138D}.
This solution is also quite different from the Riemann solution between
the shock front and contact discontinuity.

When we solve equation (\ref{crenergy}) instead of equation (\ref{crnumber}),
we cannot reproduce the Riemann solution.   The difference comes in part
from the MUSCL approach.  Fig.~\ref{CRHD2nd} is the same as 
Fig.~\ref{CRHD1st} except that the solution is of the first order accuracy 
in space and time.  All the solutions of the first order accuracy are very diffusive around the
contact discontinuity as expected.  

The Roe-type approximate 
Riemann solver is, however, still close to the Riemann solution.   
The HLL scheme 
produces a hump in $ P _{\text{g}} $ and a dump in $  P _{\text{cr}} $ 
around the contact discontinuity in the first order solution.  The hump
and dump are much smaller in the second order solution.  This implies
that they are due to numerical diffusion.  We will discuss the origin of the hump
and bump in Section 3.3.  The numerical solutions are
still acceptable in the first order accuracy when we solve the fully 
conservation form.  

The solution obtained with the vdP scheme shows serious enhancement
in $ \chi $ at the shock front.  The entropy has a dip
at the shock front and the shock propagation speed is underestimated.
The CR pressure is overestimated while the gas pressure is underestimated.
The vdP scheme diffuses the pressure balance mode seriously.   
This is because the source term diverges at the pressure balance mode 
in the vdP scheme.   
   
Also the Pdv scheme overestimates the CR pressure and underestimates
the gas pressure.   However the difference from the Riemann solution is
smaller in the solution of the first order accuracy than in that of the 
second order accuracy.   The solution of the second order 
accuracy depends a little on the choice of variables to be interpolated.  
However, we could not find any good  solution of the second
order accuracy when we solve equation (\ref{crnumber}).   
The source term is extremely large and formally 
diverges at the shock front in the Pdv scheme.   Accordingly 
the Rankine-Hugoniot relation is not satisfied. 
The numerical solution around the shock is not improved by the
MUSCL approach.

%-- Subsection 3.2 ------------------------------------------------------------------------------------------------
\subsection{Linear Wave Test}

We examine the propagation of 1D plane waves of a small amplitude in order to confirm
the accuracy of our scheme.   In the following we follow a sound wave, entropy wave,
and the pressure balance mode.  The initial state at $ t = 0 $ is taken to be
\begin{equation}
\rho = \rho _0 + \varepsilon _\rho \cos \left( \frac{2 \pi x}{\lambda} \right) ,
\end{equation}
\begin{equation}
P _{\text{g}} = P _{\text{g},0} + \varepsilon _{\rm g} \cos \left( \frac{2 \pi x}{\lambda}  \right) , 
\end{equation}
\begin{equation}
P _{\text{cr}} = P _{\text{cr},0} + \varepsilon _{\rm cr} \cos \left( \frac{2 \pi x}{\lambda}  \right) , 
\end{equation}
\begin{equation}
v _x = v _{x,0} + \varepsilon _{v} \cos \left( \frac{2 \pi x}{\lambda}  \right) ,
\end{equation}
where $ \lambda $ denotes the wavelength.  The other parameters are summarized 
in Table \ref{waveP}.   Note that the sound speed is set to be $ a = 1.0 $ in all the
models.   The time step is taken to be $ \Delta t =  0.5 \Delta x / a $.  The size
of the computation box is the same as the wavelength and the periodic boundary
condition is applied.

\begin{table}  %%%% Subsection 3.2 table2 %%%%%%%%%%%%%%%%%%%%
\caption{The initial states in the linear wave propagation tests  \label{waveP}}
\begin{center}
\begin{tabular}{lccc}
\hline
 & Sound & Entropy & Pressure \\
& & & balance \\
\hline
$ \rho _0 $ & 1.0 & 1.0 & 1.0 \\
$ P _{\text{g},0} $ & 1/3 & 1/3 & 1/3\\
$ P _{\text{cr},0} $ & 1/3 & 1/3 & 1/3 \\
$ v _{x,0} $ & 0.0 & 0.5 & 0.5 \\
\hline
$ \varepsilon _\rho ~$ ($10^{-6}$) & 1.0 & 1.0 & 0.0\\
$ \varepsilon _{\text{g}} ~$ ($10^{-6}$) & $\gamma_{\text{g}}$/3 & 0.0 & -1.0 \\
$ \varepsilon _{\text{cr}} $ ($10^{-6}$) & $\gamma_{\text{cr}}$/3 & 0.0 & 1.0 \\
$ \varepsilon _{v} ~$ ($10^{-6}$) & 1.0 & 0.0 & 0.0 \\
\hline
\end{tabular}
\end{center}
\end{table}%%%%%%%%%%%%%%%%%%%%%%%%%%%%%%%%%%%%%%

\begin{figure}   %%% Subsection 3.2 ---  Fig 5 %%%%%%%%%%%%%
\begin{center}
\includegraphics[width=1.01\columnwidth]{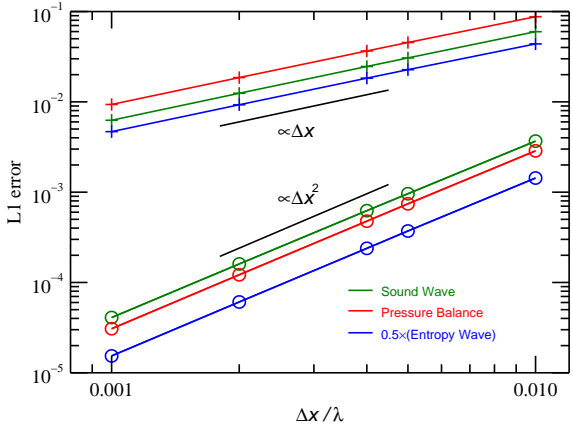}
\end{center}
\caption{\label{CRHD_acc} 
The L1 norm of the numerical error is shown as a function of the cell width 
in unit of the wavelength.  The crosses denote the errors in the solution of
the first order accuracy,  while the circles denote those of the second order
accuracy.   The green, blue and red symbols are for the sound wave, pressure
balance mode, and the entropy wave, respectively.}
\end{figure}    %%%%%%%%%%%%%%%%%%%%%%%%%%%%%%%

We measure the numerical error at the epoch at which the wave propagates by a wavelength,
i.e., $ t = \lambda / a $ for the sound wave and $ t = \lambda / v _0 $ for the other
waves.   We use the L1 norm defined as
\begin{equation}
\delta q (t) = \frac{1}{N \varepsilon _q} \sum _j 
\left| q _{\text{numerical}} (x _j, t) - q _{\text{exact}} ( x _j, t) \right| ,
\end{equation}
where the symbol, $ q $, denotes a physical quantity.   The numerical and
exact solutions are denoted by
$ q _{\text{numerical}} (x _j, t) $ and $ q _{\text{exact}} (x _j, t) $, respectively.   
The symbols, $ N $ and $ \varepsilon _q $, denote the number of cell in a wavelength
and the wave amplitude in the variable, $ q $.

Fig. \ref{CRHD_acc} denotes $ \delta \rho (\lambda/a) $ in the sound wave 
as a function of $ \Delta x / \lambda \equiv 1/N $ by the green crosses and circles,
the former and latter of which are obtained with the first- and second-order schemes,
respectively.   The red crosses and circles are the same as the green ones but
for $ \delta P _{\text{cr}} \left(\lambda/v _{x,0} \right) $ in the pressure balance mode.
We denote the error in the entropy mode, $ \delta \rho \left(\lambda /v _{x,0}\right)/2 $,
by the blue crosses and circles.   The factor, 1/2, is introduced to avoid the overlap with
the red symbols.   Our second order schemes solve all these hydrodynamical waves 
with the second order accuracy in space.

Also the Pdv and vdP scheme provide the solutions of the second
order accuracy for the test problems examined in this subsection. As far as
the wave amplitude is small, all the schemes provide nearly the same results
as expected. 

%---- Subsection 3.3 -------------------------------------------------------------------------------------- %
\subsection{Advection of the Pressure Balance Mode}

The pressure balance mode produces some unphysical features such as 
spurious enhancement in the gas pressure and dent in the CR pressure
as shown in Fig.~\ref{CRHD1st}.  It also emits spurious sound waves as
shown later in this subsection.  We identify the origin and 
provide a remedy. 
The initial state of the test problem is summarized 
in Table \ref{advPBtable}.  The velocity should remain constant in this 
problem since the total pressure has the same value in the both sides.   
The cell width and time step are taken to be  $ \Delta x = 0.1 $ 
and $ \Delta t = 0.025 $, respectively. 
Accordingly the CFL number is 0.585 in the simulations shown below.

\begin{figure*}  %%%  Subsection 3.3 --- Fig 6 %%%%%%%%%% 
\begin{center}
\includegraphics[width=0.9\textwidth]{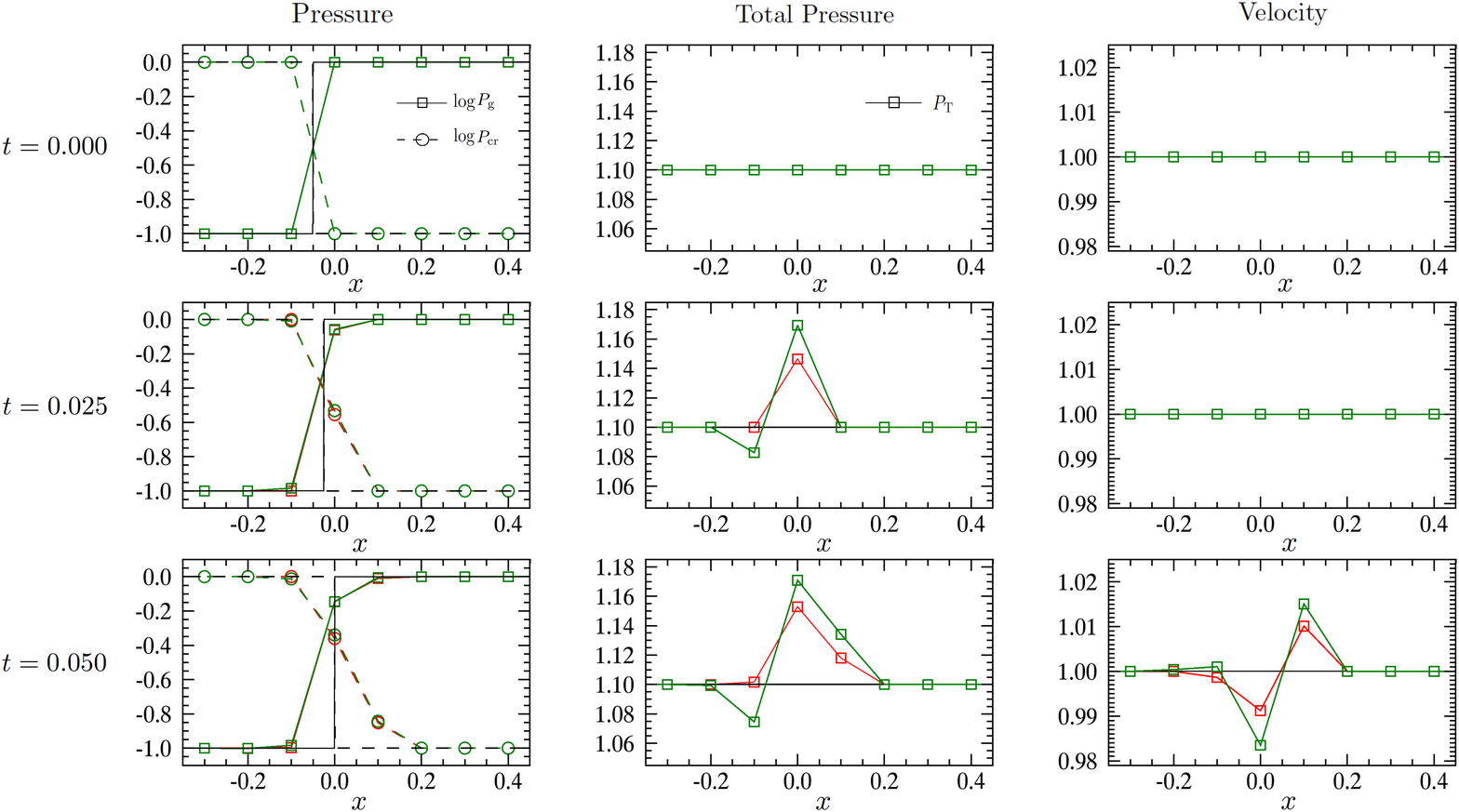}
\end{center}
\caption{\label{CRHDadvection}  
Test for the advection of the pressure balance mode.   The top panels
denote the gas pressure, the CR pressure, the total pressure, and the velocity 
in the initial state from left to right.  The middle and bottom panels denote  
those at the first time step, $ t = \Delta t $, and at the second time step, 
$ t =  2 \Delta t $, respectively.
The green and red curves denote the solutions obtained with HLL
and Roe-type numerical fluxes, respectively.  The black line
denotes the exact solution.}
\end{figure*}

The spurious sound wave appears from the beginning irrespectively of the
numerical scheme applied.  Fig.~\ref{CRHDadvection} shows the solutions 
at the first three time steps, i.e., at $ t = 0.000 $, 0.025, and 0.050  from top to bottom.  
The left-hand and central panels show $ P _{\text{g}} $ and
$ P _{\text{T}} $, respectively, while the right-hand panels show $ v _x $.
The green and red curves with the symbols denote the solutions obtained
with the HLL scheme and the Roe-type Riemann solver, respectively, while
the black lines denote the exact solution.   The accuracy of these solutions
is the first order in space and in time.   The total pressure is enhanced
at the second time step ($ t = 0.025 $) and the velocity perturbation appears 
at the third time step ($ t = 0.050 $).

\begin{figure*}  %%%  Subsection 3.3 --- Fig 6 %%%%%%%%%% 
\begin{center}
\includegraphics[width=0.9\textwidth]{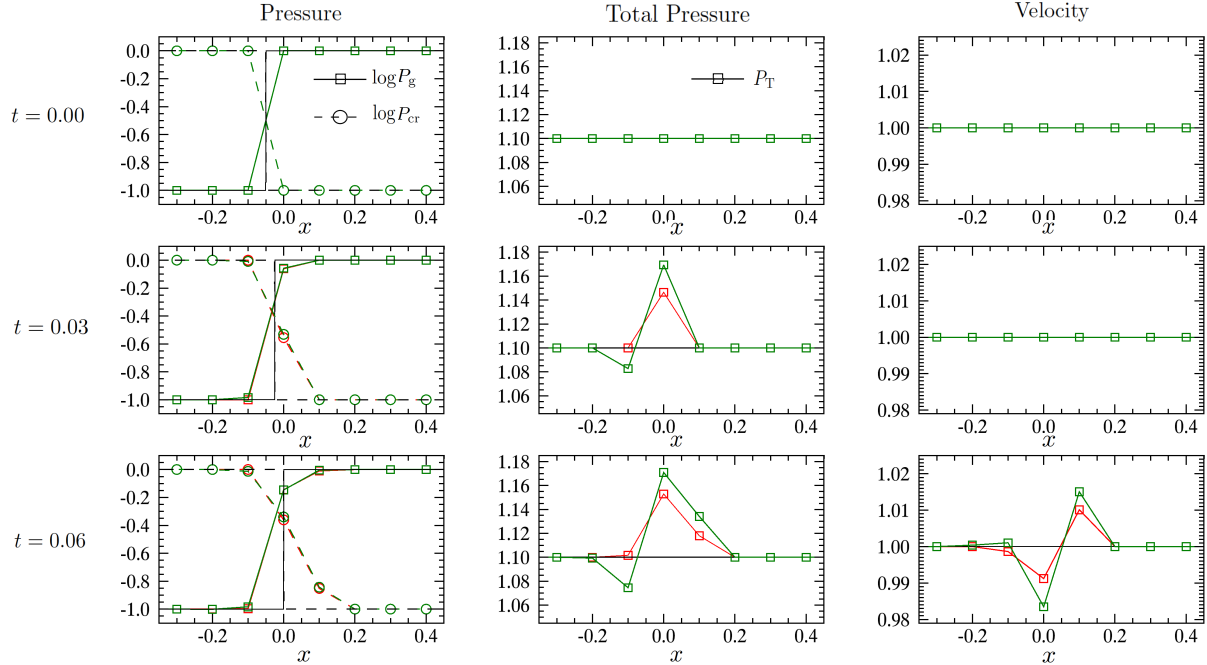}
\end{center}
\caption{\label{CRHDadvection_2nd}  
The same as Fig.~\ref{CRHDadvection} but for the 2nd order accuracy. 
}
\end{figure*}

Fig.~\ref{CRHDadvection_2nd} is the same as Fig.~\ref{CRHDadvection} but for the solutions of
the second order accuracy.   The velocity perturbation appears from the
first step, $ t = 0.025 $.   The amplitude of the perturbation is nearly
the same as that in the solutions of the first order accuracy.   As shown later, the 
perturbation can suppressed if the initial state is slightly modified.

\begin{figure}   %% Subsection 3.3 -- Fig. 7 %%%%%%%%%%
\centering
\begin{center}
\includegraphics[width=\columnwidth]{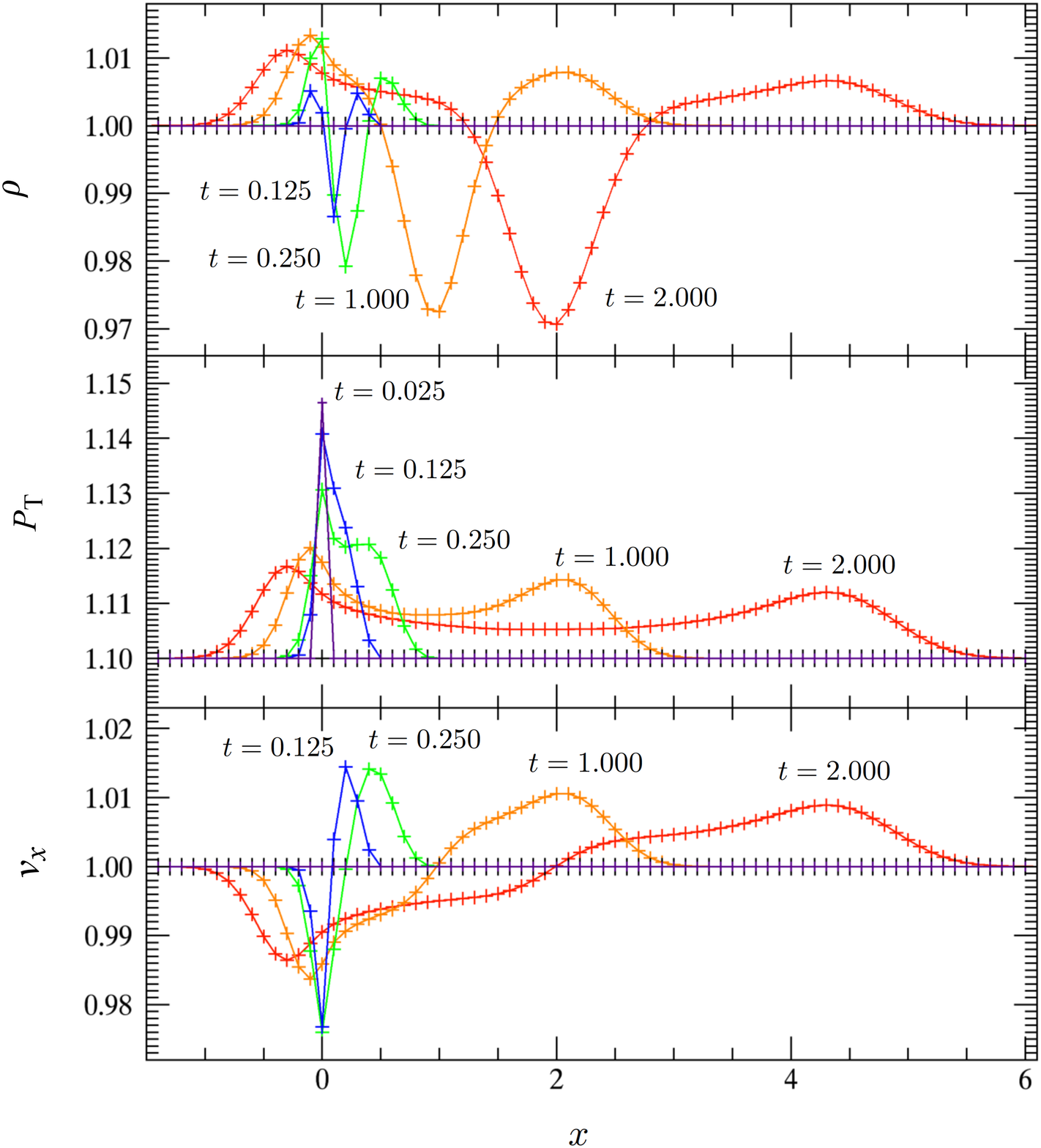}
\end{center}
\caption{\label{numericaldif} 
Late stages of the test problem shown in Fig. 6
The panels denote $ \rho $, $ P _{\text{T}} $, and $ v _x $ 
from the top to the bottom.   The symbols and curves denote
the solution at $ t = 0.025 $, 0.125, 0.250, 1.0, and 2.0.
}
\end{figure}

Fig.~\ref{numericaldif} shows the later stages of the solutions obtained
with the first order Roe scheme.  The top, middle and bottom panels
show the density, the total pressure and the velocity, respectively.
The colour denotes the epoch.   The velocity perturbation observed in
Fig.~\ref{CRHDadvection} evolves into sound waves propagating 
rightward and leftward of which phase velocity are 
$ v _x + a = 2.342 $ and $ v _x - a = -0.225 $  for the former 
and latter, respectively.   In addition to the sound waves, the density
profile has a dent around the contact discontinuity.   The dent grows
in depth and in width.   We need to suppress the sound waves of
the numerical origin and growth of the dent.

\begin{figure}    %%% Subsection 3.3 -- Fig. 8  %%%%%%%% 
\centering
\begin{center}
\includegraphics[width=\columnwidth]{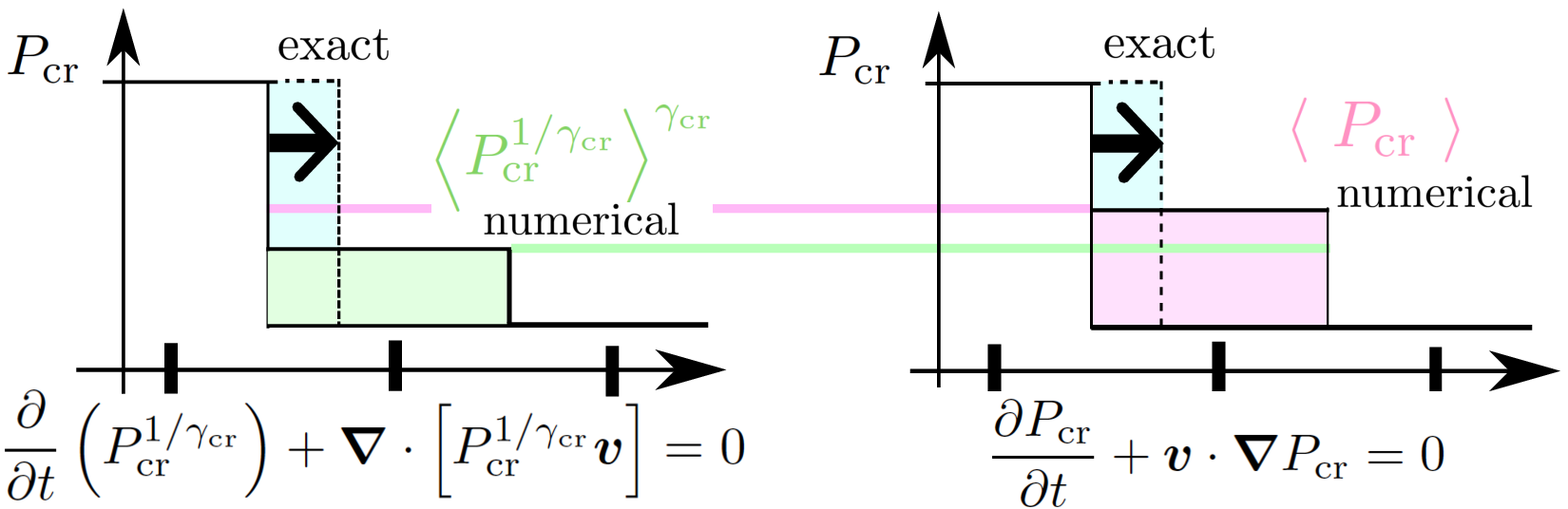}
\end{center}
\caption{\label{numericaldif2} Illustration of the mechanism for the decrease 
in $ P _{\text{cr}} $ due to numerical diffusion by advection of the pressure
balance mode.  The left-hand panel denotes the numerical solution of the CR number 
conservation, while the right does that of the advection equation.}
\end{figure}

This artificial wave excitation is due to the numerical diffusion of CR numbers.
If we take account of numerical diffusion, equation (\ref{crnumber}) is
rewritten as
\begin{equation}
\frac{\partial}{\partial t} \rho _{\text{cr}} + \bm{\nabla} \cdot \left( \rho _{\text{cr}}
\bm{v} \right) = \bm{\nabla} \cdot \left( \eta \bm{\nabla} \rho _{\text{cr}} \right) , \label{crnumbermod}
\end{equation}
where $ \eta $ denotes the effective diffusion coefficient.   Hence equation 
(\ref{crenergy}) is rewritten as
\begin{equation}
\begin{split}
%\frac{\partial}{\partial t}  \left( \frac{ P _{\text{cr}} }{ \gamma_{\text{cr}} -1 }  \right) + 
%\bm{\nabla} \cdot \left[ \frac{ \gamma_{\text{cr}} }{\gamma_{\text{cr}}-1 }  P _{\text{cr}} \bm{v} - \eta \bm{\nabla} \left( \frac{ P _{\text{cr}} }{ \gamma_{\text{cr}} -1 } \right) \right] \nonumber \\
%= \bm{v} \cdot \bm{\nabla} P_{\text{cr}} - \frac{\eta}{\gamma _{\text{cr}} P _{\text{cr}}}
% \left| \bm{\nabla} P _{\text{cr}} \right| ^2 , \label{Pcrdif}
\frac{\partial}{\partial t}  \left( E_{\text{cr}}  \right) + 
\bm{\nabla} \cdot \left[  \left( E_{\text{cr}} + P_{\text{cr}} \right) \bm{v} - \eta \bm{\nabla} 
\left( E _{\text{cr}} \right) \right]  \\ 
= \bm{v} \cdot \bm{\nabla} P_{\text{cr}} - \frac{\eta}{\gamma _{\text{cr}} P _{\text{cr}}}
 \left| \bm{\nabla} P _{\text{cr}} \right| ^2 , \label{Pcrdif} 
\end{split}
\end{equation}
where the second term in the right hand side denotes the net loss in the CR energy.
Fig. \ref{numericaldif2} illustrates this mechanism of the loss in the CR energy

\begin{table}   %%% Susection 3.3 --- Table 2 %%%%%%%%
\caption{\label{advPBtable} The initial state in the test problem of  the advection of pressure balance mode.}
\begin{center}
\begin{tabular}{rcccc}
\hline
~~ & $\rho$ & $v_x $ & $P_{\text{g}}$ & $P_{\text{cr}}$ \\ \hline
Left ($ x < 0 $) & 1.0 & 1.0 &  0.1 & 1.0 \\ 
Right ($ x \ge 0 $) & 1.0 & 1.0 &  1.0 & 0.1 \\ \hline
\end{tabular}
\end{center}
\end{table}        %%%%%%%%%%%%%%%%%%%%%%%%%%%

   %%%%%%%%%%%%%%%%%%%%%%%%%%%%

   %%%%%%%%%%%%%%%%%%%%%%%%%%%%

    %%%%%%%%%%%%%%%%%%%%%%%%%%%%

      %%%%%%%%%%%%%%%%%%%%%%%%%%%%

The loss in the CR energy is compensated by the gain 
in the gas energy, since the total energy is conserved.   
We can evaluate the change in the total pressure as follows.  Given the CR pressure
decrease by $\Delta P _{\text{cr}} $, the CR energy decreases by
\begin{equation}
\Delta E _{\text{cr}} = \frac{\Delta P _{\text{cr}}}{\gamma _{\text{cr}} - 1} ,
\end{equation}
and the gas energy increases by the same amount.   Hence the gas pressure
increases by
\begin{equation}
\Delta P _{\text{g}} = -  \left( {\gamma _{\text{g}} - 1} \right) \Delta E _{\text{cr}} 
 = - \frac{ \gamma _{\text{g}} - 1}{\gamma _{\text{cr}} - 1 } \Delta P _{\text{cr}} .
\end{equation}
Then the total pressure increases by
\begin{equation}
\Delta P _{\text{T}} = \frac{\gamma _{\text{cr}} - \gamma _{\text{g}}}
{\gamma _{\text{cr}} - 1} \Delta P _{\text{cr}} ,  \label{PTdif}
\end{equation}
since $ \gamma _{\text{cr}} < \gamma _{\text{g}} $ and $ \Delta P _{\text{cr}} < 0 $.

The emission of the sound waves result in the decrease in the density.
Thus, also the dent is due to the numerical diffusion of the CR pressure.

\begin{figure}   %% Subsection 3.3 -- Fig. 7 %%%%%%%%%%
\centering
\begin{center}
\includegraphics[width=\columnwidth]{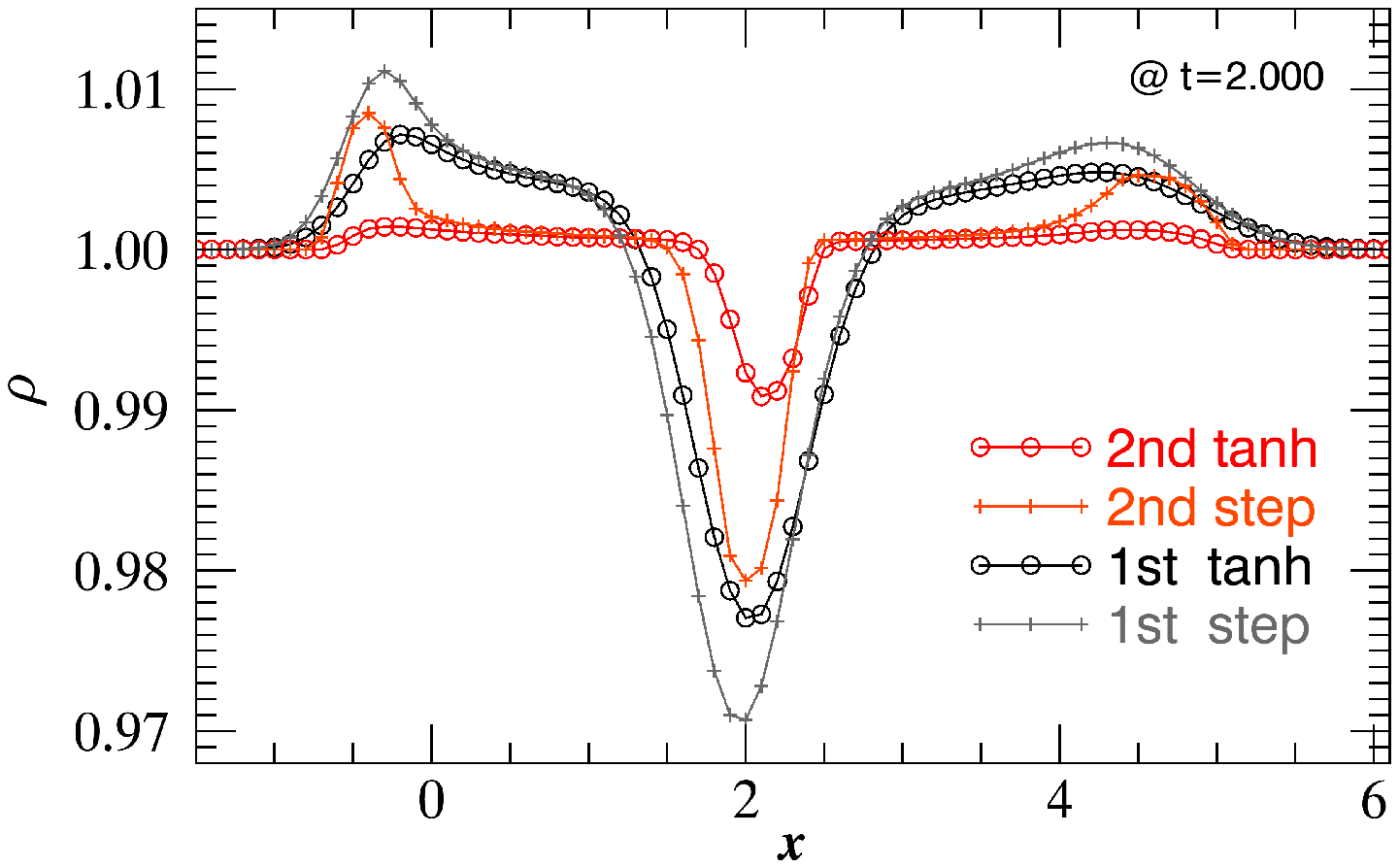}
\end{center}
\caption{\label{advPB_all}
Compared are the numerical solutions of the advection test problem at $ t = 2.0 $.
The initial state is expressed by equation (78) in the solutions denoted by red and black,
while it is by the step function in the solutions denoted by orange and grey.
The red and orange curves denote the solutions of the second order accuracy,
while the black and grey do those of the first order accuracy.
All the solutions are obtained with the Roe-type Riemann solver.  
}
\end{figure}    %%%%%%%%%%%%%%%%%%%%%%%%%%%%

Although the numerical diffusion is inevitable,  the effects can be alleviated 
by modifying the initial transition less sharp.  If the initial profile is modified
to be
\begin{equation}
q (x) = \frac{q _{\rm R} + q _{\rm L}}{2} +
\frac{q _{\rm R} - q _{\rm L}}{2} \tanh \left( \frac{x}{h} \right) , \label{tanh}
\end{equation}
where $ q $ and $ h $ denote a physical variable and the thickness of the 
transition layer, respectively.  The subscripts, R and L, 
denote the values at the right and left states, respectively.   
Fig.~\ref{advPB_all}  shows the effects of the modified initial profile.   All the curves
denote the solutions of the advection test at $ t = 2.000 $.  The red and
orange curves denote the solutions of which initial profile is given by 
equation (\ref{tanh}) with $ h = \Delta x = 0.1 $.   The red solution is obtained
with the second order Roe scheme, while the orange is with the first
order Roe scheme.  The black and grey curves denote the solutions of which
initial state is denoted by a step function ($ h = 0 $).  The black and grey
solutions are obtained with the second and first order Roe schemes, respectively.
Adoption of a second order scheme alone cannot suppress the numerical 
diffusion effectively.  Only when $ h \ga \Delta x $, the numerical diffusion
is suppressed sufficiently.

The Pdv scheme does not produce the spurious wave, since 
the source term vanishes in this test problem.  The vdP scheme suffers
from the spurious wave since the source term diverges at the front of the
pressure balance mode.   We have not tried to improve the vdP scheme
to suppress the spurious wave excitation, since the vdP scheme can not
solve a shock wave properly.

We encounter a similar problem when
solving the advection of the tangential shear.  The shear velocity is smeared by
numerical diffusion and hence a part of the kinetic energy is lost.  The loss
is compensated by the increase in the internal energy and hence the gas 
pressure increases spuriously around the shear.  This problem is also 
alleviated by the same recipe.

%= Subsection 3.4===============================%
\subsection{1D MHD shock tube test}

In this subsection we demonstrate that the 1D MHD shock tube problem can
be solved with the Roe-type approximate Riemann solver.
The initial state of the problem is summarized in Table \ref{exMHD}.
This problem is similar to the well-known one tested by \cite{1988JCoPh..75..400B}.
The resolution, the time step, and CFL number are taken to be 
$ \Delta x = 1/256 $, $ \Delta t = 8 \times 10^{-4} $, and 0.843, respectively.

\begin{figure*}   %%%% Subsection 3.4 - Fig. 9 %%%%%%%%
\begin{center}
\includegraphics[width=\textwidth]{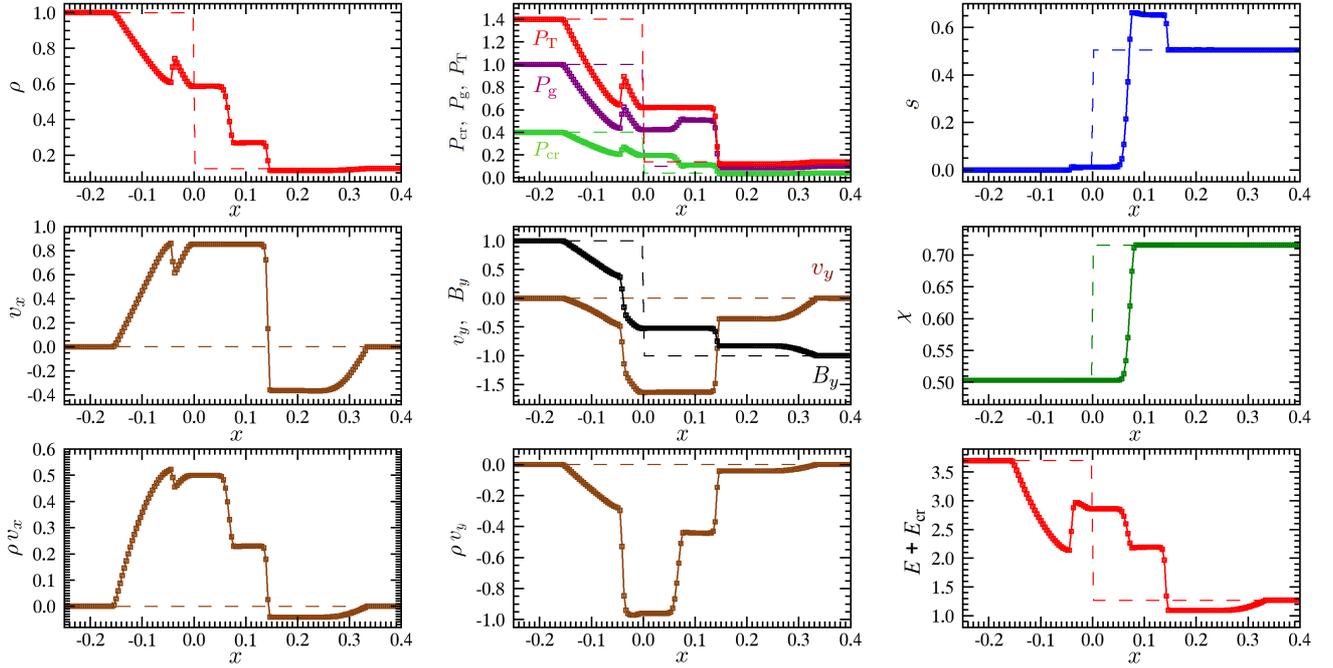}
\caption{\label{CRMHD} 
An example of CR MHD shock tube problem.  Top: 
the density, CR pressure, gas pressure, total pressure, and
entropy from left to right.    Middle: the velocity, magnetic 
field, and CR concentration from left to right.
Bottom: the $ x $- and $ y $-components of the momentum density, and the total energy
density from left to right.}
\end{center}
\end{figure*}

Fig. \ref{CRMHD} shows the numerical solution at $ t = 0.08 $.
The left-hand panels denote $ \rho $, $ v _x $, and $\rho v _x $, respectively.
The solid curves with the open squares denote the numerical solution while
the dashed ones do the initial state.  The top central panel denotes
$ P _{\text{T}} $ (red), $ P _{\text{g}} $ (purple), and $ P _{\text{cr}} $
(yellow green), while the middle central one does $ v _y $ 
(brown) and $ B _y $ (black), and the bottom central one dose $ \rho v _y$.   
The right-hand panels denote $ s $, $ \chi $,  and $ E + E _{\text{cr}} $, respectively.

     %%%%%%%%%%%%%%%%%%%%%%%%%%%

The solution shows the fast rarefaction wave, the slow shock, the contact 
discontinuity (the pressure balance mode), the slow compound, and the
fast rarefaction, from right to left.   Note that the contact discontinuity and
the pressure balance mode are solved sharply.   The CR concentration, 
$ \chi $, changes only at the contact discontinuity. 

\begin{table}    %%%% Subsection 3.4 - Table 4 %%%%
\caption{\label{exMHD}
Initial left- and right-state in the shock tube problem of modified \protect \cite{1988JCoPh..75..400B}. }
\begin{center}
\begin{tabular}{rlllllllll}
\hline
& $\rho$ & $ B _x $ & $B_y$ & $B_z$ & $P_{\text{g}}$ & $P_{\text{cr}}$ \\ \hline
Left $(x < 0)$ & 1.0  &1.0 & ~1.0 & 0.0 & 1.0 & 0.4 \\ 
Right $(x\geq 0)$ & 0.125  & 1.0 & -1.0 & 0.0 & 0.1 & 0.04 \\ \hline
\end{tabular}
\end{center}
\end{table}      %%%%%%%%%%%%%%%%%%%%%%%

%==3.5==========================================% 
\subsection{2D MHD expansion}

\begin{table}  %%% Subsection 3.5 - Table 5  %%%%%%%%%
\caption{\label{blastwave}Initial inside- and outside-state in the 2D MHD expansion. }
\begin{center}
\begin{tabular}{rlllllllll}
\hline
& $\rho$ &  $ B _r $ & $B_{\varphi}$ & $B_z$ & $P_{\text{g}}$ & $P_{\text{cr}}$ \\ \hline
Inside    & 1.0 & 0.0 & 0.0 & 0.4472 & 2.0 & 1.0 \\ 
Outside  & 0.2 & 0.0 & 0.0 & 0.4472 & 0.02 & 0.1 \\ \hline
\end{tabular}
\end{center}
\end{table}   %%%%%%%%%%%%%%%%%%%%%%%%%%%

\begin{figure*}  %%% Subsection 3.5 - Fig.  %%%%%%%
\begin{center}
\includegraphics[width=\textwidth]{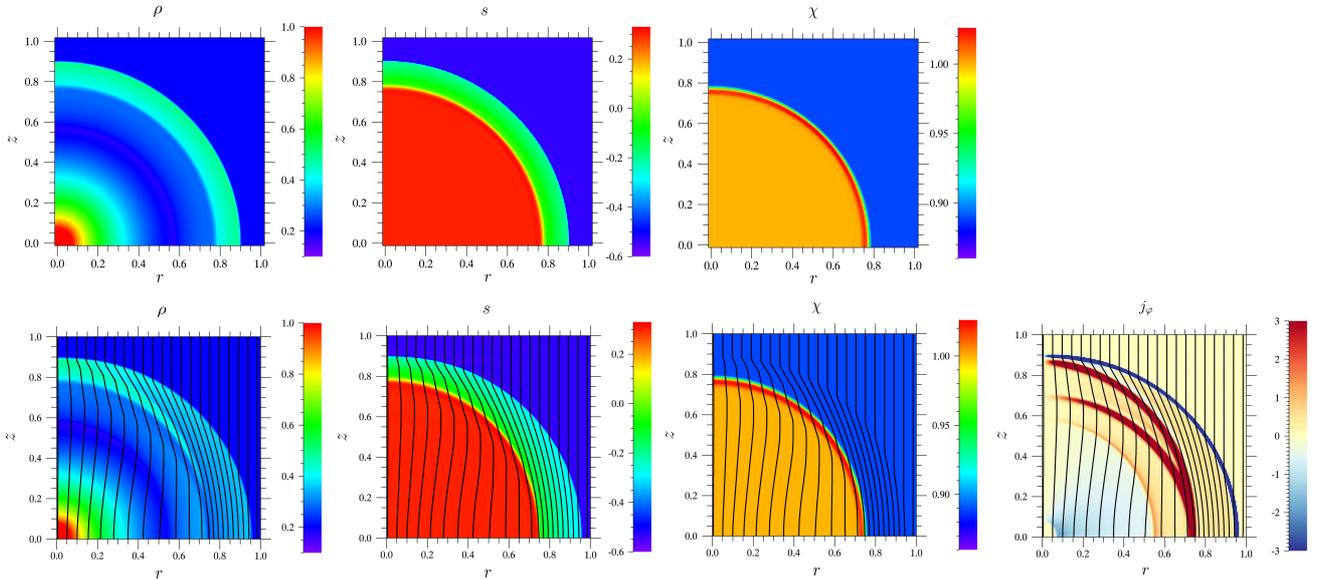}
\end{center}
\caption{\label{bw2d}  Expansion of a high pressure sphere.
Top:  the density, entropy, and CR concentration in HD expansion
problem from left to right.  Bottom: the density, entropy, CR concentration,
and electric current density in MHD problem.
The initial states are the same except for the initial magnetic field.
The black curves denote the magnetic field in the bottom panels.}
\end{figure*}    %%%%%%%%%%%%%%%%%%%%%%%%%%

In this subsection we consider an expansion of a hot sphere. 
 We use the cylindrical coordinates, 
$ (r, \varphi , z ) $, and assume the symmetry around the $ z $-axis.
 The gas pressure is
higher inside the sphere of $ \sqrt{r ^2 + z ^2}  = 0.5 $ in the initial state.
It is assumed to have the profile,
\begin{equation}
\begin{split}
 P _{\text{g}} (r,z) = P_{\text{g,in}} &+ \frac{(P_{\text{g,out}}- P_{\text{g,in}})}{2} \\
 &~~~~\times \left[ \tanh \left(\frac{ \sqrt{r^2+z^2} -0.5 }{ 0.01 } \right) +1 \right] ,
\end{split}
\end{equation}
where the subscripts, in and out, denote the values inside and outside
the sphere, $ \sqrt{r ^2 + z ^2} = 0.5 $.
Also the CR pressure and density are assumed to have similar profiles.
The parameters are summarized in Table \ref{blastwave}.  
The velocity is assumed to vanish in the initial state.
The spatial resolution, time step, and CFL number are taken to be
$ \Delta x = \Delta z =1/200$, $\Delta t = 5 \times 10^{-4} $, and 0.340
respectively. 

We construct two models having the same initial $\rho $, $ P _{\text{g}} $,
and $ P _{\text{cr}} $.   The magnetic field vanishes in the first model while
it is uniform and has only the $ z $-component in the second model.
The plasma beta is $ \beta = 20 $ inside and 2 outside the sphere in the
second model.  We solved this model with the Roe-type approximate
Riemann solver.

Fig. \ref{bw2d} denotes the stage at $ t = 0.19 $.  The top panels denote
$ \rho $, $ s $, and $ \chi $ in the first model from left to 
right.   The values are shown by colour and the colour bars are shown in the
right of each panel.
 They show spherical shock wave propagating outward and rarefaction
propagating inward.  The contact discontinuity and pressure balance
mode are clearly resolved.

The bottom panels denote $ \rho $, $ s $, $\chi $,
and the electric current density, $ j _{\varphi} $.   The colour denotes the values while the black
contours do the magnetic field lines.   The expansion is aspherical
due to the magnetic field.  The model shows three shock waves:
a fast shock wave and two slow shock waves.  The magnetic field
is bent sharply at the shock fronts.  Note that the electric current
density is confined at the shock fronts and in the rarefaction wave.
This example demonstrates that the multi-dimensional CR MHD
equations can be solved successfully if we employ the fully
conservation form.

%% Section 4   %%%%%%%%%%%%%%%%%%%%%%%%%%%%%%%%%% 
\section{Discussions and Conclusions}

We have succeeded in rewriting the CR MHD equations in the fully
conservation form as shown in Section 2.  We discuss the advantages of
the fully conservation form in this section.

First the Rankine-Hugoniot relation is automatically fulfilled when 
the CR MHD equations are integrated in the fully conservation form.
Thus, the jumps in the density and pressure at MHD shocks are
evaluated correctly in the solutions.  When we integrate the CR MHD
equations in the original form, the numerical solution may violate the 
Rankine-Hugoniot relation as shown in Section 3.3.  

One might think that any shock tube problems could be solved 
without using the fully conservation form.
\cite{2016A&A...585A.138D}  have succeeded
in solving a CR shock tube problem posed by \cite{2006MNRAS.367..113P} 
by evaluating Pdv as the source term.
However, the success is in part due to the fact that CR pressure is 
smaller than the gas pressure in the post-shocked gas.  
Table~\ref{pfrommer06} summarizes the pressure and density distributions
in the test problem.  The number in the first column species the intervals
where the density and pressure are constant in the Riemann solution
(see Fig. \ref{CRHDshock}).  The values in regions 2 and 3 are 
given by the Riemann solution and match their Fig. 3.
The CR pressure is dominant in regions 3 and 5 while it is not in region 2, 
i.e., in the post-shocked gas.
Then the source term has a minor contribution at the shock front.
See Appendix B for more details on these test problems. 

\begin{table}     %%%  Section 4 -  Table 5  %%%%%%%%
\caption{The shock tube problem of \protect \cite{2016A&A...585A.138D}. 
The adiabatic indexes are taken to be $ \gamma _{\text{g}} = \gamma_{\text{cr}} = 1.4$
in this test problem.
}
\label{dubois15}
\begin{center}
\begin{tabular}{cccc}
\hline
Region & $\rho$ &  $ P _{\text{g}} $ & $ P _{\text{cr}} $ \\
\hline
1 & 0.100 & 0.066 & 0.034  \\
2 & 0.204 & 0.192 & 0.093 \\
3 & 0.408 & 0.097 & 0.187 \\
5 & 1.000 & 0.340 & 0.660  \\
\hline
\end{tabular}
\end{center}
\end{table}   %%%%%%%%%%%%%%%%%%%%%%%

\cite{2006MNRAS.367..113P} also succeeded in solving a shock tube problem
with their smoothed particle hydrodynamics (SPH) code, although their solution shows small oscillations
around shock front and contact discontinuity.   However, the CR pressure
is much smaller than the gas pressure in the post-shocked gas.
Table~\ref{pfrommer06} is the same as Table~\ref{dubois15} but for
the shock tube problem shown in Fig.~1 of \cite{2006MNRAS.367..113P}.

\begin{table}  %%% Section 4 -- Table 6 %%%%%%%%
\caption{The shock tube problem of \protect \cite{2006MNRAS.367..113P}. }
\label{pfrommer06}
\begin{center}
\begin{tabular}{cccc}
\hline
Region & $\rho$ &  $ P _{\text{g}} ~ (\times 10^{4}) $ & $ P _{\text{cr}} ~ (\times 10^4)$ \\
\hline
1 & 0.200 & $0.024$ & $~0.024$ \\
2 & 0.796 & $5.141$ & $~0.147$ \\
3 & 0.400 & $1.455$ & $~3.832$ \\
5 & 1.000 & $6.700$ & $13.000$  \\
\hline
\end{tabular}
\end{center}
\end{table}   %%%%%%%%%%%%%%%%%%%%%
 
The violation of the Rankine-Hugoniot relation is serious 
when the CR pressure dominates in the post-shocked gas.
A clear example is shown in Section 3.3.

The fully conservation form has another advantage, adaptation
to higher order scheme.   Various standard higher order schemes
 are available, when differential 
equations are written in the fully conservation form.
see e.g. \citealt{toro2009riemann} for the methods to achieve higher order
accuracy.   Remember that the source terms and numerical fluxes 
have been evaluated separately.  Thus the source term can be another
source of numerical oscillation when it is evaluated to be of higher
order accuracy in space.

The fully conservation form may be useful when we take account of
injection of CRs.  CRs can be generated from supernova explosions and 
diffusive shock acceleration.   The generation can be taken into account
in the CR MHD equations, if it is modelled successfully (see e.g. \citealt{1993ApJ...406...67Z};
\citealt{2008A&A...481...33J}; \citealt{2012MNRAS.421.3375V}).
Thus far, only the energy injection rate, is taken into account in the literature.
However, we can take account of both the energy injection rate, $ S _{E} $,
and the injection rate in number, $ S _{\rho} $, in the fully conservation
form.  In the following we assume that $ E _{\text{cr}} $ and $ \rho _{\text{cr}} $
are related by
\begin{equation}
E _{\rm cr} = \frac{K \rho _{\text{cr}} ^{\gamma _{\text{cr}}}}{\gamma _{\text{cr}} - 1} ,
\end{equation}
where $ K $ denotes the CR {\it entropy} and has been assumed to be $ K = 1 $
thus far.  If the injection is taken into account, the change in $ K $ should
be described as
\begin{eqnarray}
\frac{1}{K}\frac{dK}{dt} & = & 
\frac{S _{E}}{E _{\text{cr}}} - \gamma _{\text{cr}} \frac{S _\rho}{\rho _{\text{cr}}} \\
& = & \frac{S _\rho}{E _{\text{cr}}} \left( \frac{S _{E}}{S _\rho} - \gamma _{\text{cr}}
\frac{E  _{\text{cr}}}{\rho _{\text{cr}}} \right) , \label{averageCR}
\end{eqnarray}
where $ dK/dt $ denotes the Lagrangian derivative.
Equation (\ref{averageCR}) follows the change the average energy of CRs, 
$ E _{\text{cr}} / \rho _{\text{cr}} $.  The average CR energy is important information
to estimate the diffusion coefficient.

This paper has proved the usefulness of CR number conservation,
equation (\ref{crnumber}).   It is derived from and equivalent to 
the CR energy equation, equation (\ref{crenergy}).  However, the former
is written in the fully conserved form, while the latter is not.
The former is more suitable for numerical analysis than the latter,
since the approximate Riemann solutions are given explicitly and 
hence shock waves are reproduced without numerical oscillations.
The derived approximate Riemann solutions are only slightly 
different from those for the ideal MHD equations: the sound speed 
is modified by inclusion of the CR pressure and a new mode,
the pressure balance mode, is added.  Thus we can construct
higher order scheme by applying the methods developed for the
ideal MHD equations.  We have also suggested to extend the CR MHD
equations by introducing the CR entropy.  This extension enables us
to evaluate the average CR energy.   It should be useful to estimate the
diffusion and emission from CRs.

\section*{Acknowledgements}

We thank Ryoji Matsumoto and Yosuke Matsumoto for their helpful
discussions and suggestions.  



%\bibliographystyle{mnras}
%\bibliography{16mar25} % if your bibtex file is called example.bib

%%%%%%%%%%%%%%%%%%%%%%%%%%%%%%%%%%%%%%%%%%%%%%%%%%

%%%%%%%%%%%%%%%%% APPENDICES %%%%%%%%%%%%%%%%%%%%%

\appendix

%Appendix A%%%%%%%%%%%%%%%%%%%%%%%%%%%%%%%%
\section{Roe-type Approximate Riemann Solver}

In this appendix  we derive the Roe-type approximate Riemann solver
for the CR HD equations from equations (\ref{dcomps}) through 
(\ref{dcompe}).   They show that the spatial difference is decomposed
into four waves.   When the wave speeds ($\lambda _i $) and 
eigenvectors ($ \bm{r} _i$) are evaluated by the Roe-average 
defined as
\begin{equation}
\begin{split}
\bar{\rho}& = \sqrt{\rho_{j+1} \rho_{j}}, ~~~
\bar{v}_x  = \frac{\sqrt{\rho_{j+1}} v_{x,j+1}  +\sqrt{\rho_{j}} v_{x,j} }{ \sqrt{\rho_{j+1}}  +\sqrt{\rho_{j}} },  \\
\bar{H} &=  \frac{\sqrt{\rho_{j+1}} H_{j+1}  +\sqrt{\rho_{j}} H_j }{ \sqrt{\rho_{j+1}}  +\sqrt{\rho_{j}} }, \\
\bar{P}_{\text{T}} &= \frac{\sqrt{\rho_{j+1}} P_{\text{T}, j}  +\sqrt{\rho_{j}} P_{\text{T}, j+1} }{ \sqrt{\rho_{j+1}}  +\sqrt{\rho_{j}} }     ,   \\ 
\bar{\rho}_{\text{cr}} &=\frac{\sqrt{\rho_{j+1}} \rho_{\text{cr}, j}  +\sqrt{\rho_{j}} \rho_{\text{cr}, j+1} }{ \sqrt{\rho_{j+1}}  +\sqrt{\rho_{j}} } ,  \\
\bar{a}^2  &= (\gamma_{\text{g}} -1) \left( \bar{H} - \frac{\bar{v}_x^2}{2} -\frac{ \bar{\rho}_{\text{cr}} }{ \bar{\rho} } \zeta  \right), \\
\zeta &= \frac{ \gamma_{\text{g}} - \gamma_{\text{cr}} }{(\gamma_{\text{g}} -1)(\gamma_{\text{cr}} - 1)} \frac{ P_{\text{cr},j+1} - P_{\text{cr},j} }{ \rho_{\text{cr},j+1} - \rho_{\text{cr},j}   }.
\end{split}
\end{equation}
the spatial difference is decomposed into the linear combination at the waves completely (see e.g. \citealt{toro2009riemann}). 
Here the subscripts, $ j $ and $ j + 1$, specify the cells at the centre
of which the variables are evaluated.

The Roe-type numerical flux is given by
\begin{equation}
\bm{F}_{j+1/2}^{\rm Roe} = \frac{1}{2} \left( \bm{F}_{j+1} + \bm{F}_j - 
\displaystyle\sum\limits_{i} w_i | \lambda_i | \bm{r}_i \right), \label{Roe}
\end{equation}
where the subscript $ j +1/2 $ denotes the value on the surface
between the $ j $-th and $ j +1 $-th cells.  

The Roe-type numerical flux is obtained similarity for the
CR MHD equations.  The Roe-averaged magnetic field is given by
\begin{equation}
\bar{B} _y = \frac{\sqrt{\rho _j} B _{y,j+1} + \sqrt{\rho _{j+1}} B _{y,j}}
{\sqrt{\rho _{j+1}} + \sqrt{\rho _j}} ,
\end{equation}
\begin{equation}
\bar{B} _z = \frac{\sqrt{\rho _j} B _{z,j+1} + \sqrt{\rho _{j+1}} B _{z,j}}
{\sqrt{\rho _{j+1}} + \sqrt{\rho _j}} .
\end{equation}

\begin{figure*}  %%% B - Fig. B1 %%%%%%%
\begin{center}
\includegraphics[width=0.9\textwidth]{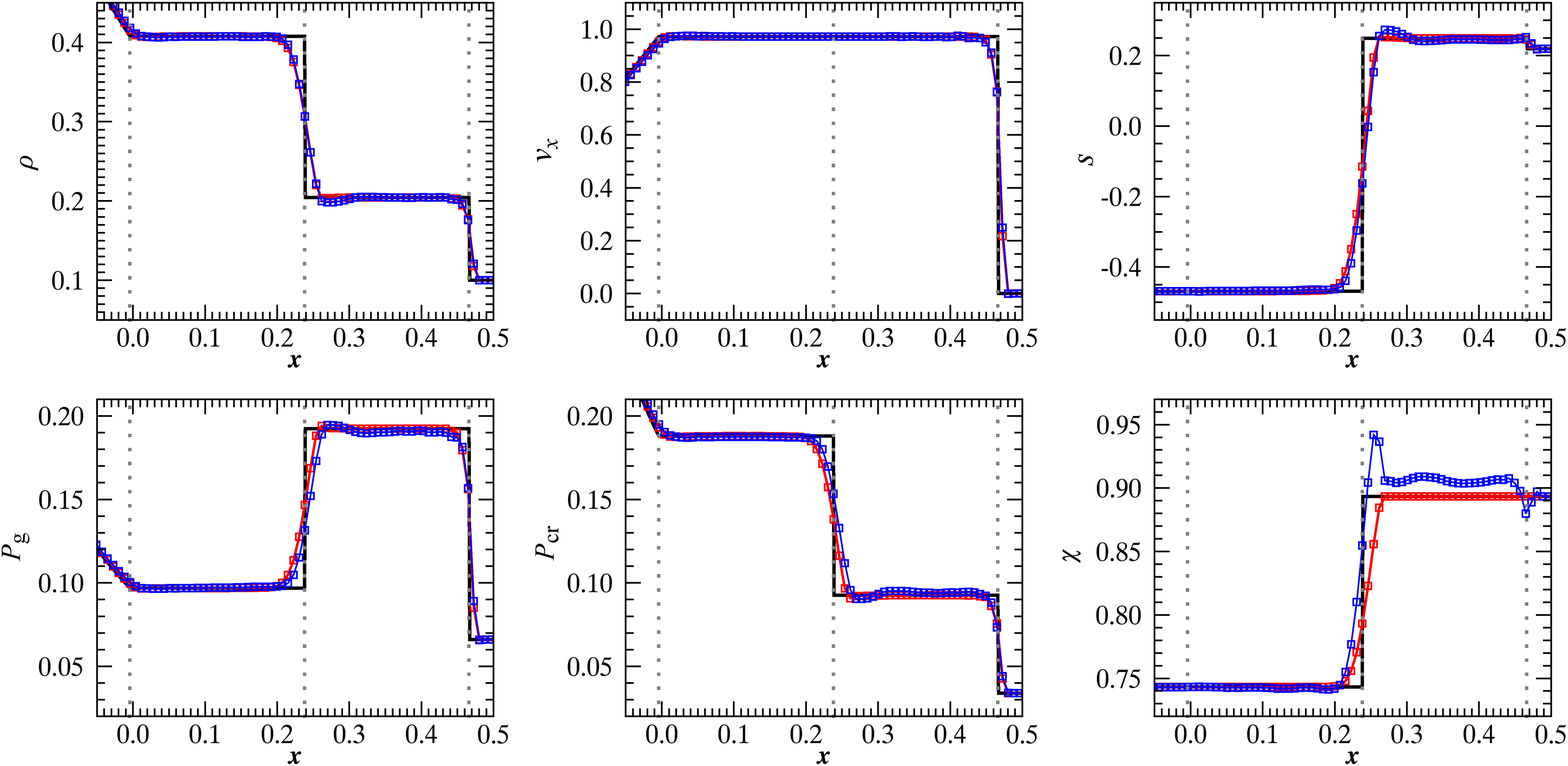}
\end{center}
\caption{\label{D15}  
The  CRHD shock tube problem of \protect \cite{2016A&A...585A.138D}.
The grey dotted lines denote the tail of rarefaction ($x= 4.218 \times 10^{-3}$), contact discontinuity 
($x= 0.2380$), and shock ($x= 0.4660$) at $ t = 0.245$  from left to right. 
 The CFL number is taken to be $ 0.743$.
}
\end{figure*}    %%%%%%%%%%%%%%%%%%%%%%%%%%

\begin{figure*}  %%%  B  - Fig. B2  %%%%%%%
\begin{center}
\includegraphics[width=0.9\textwidth]{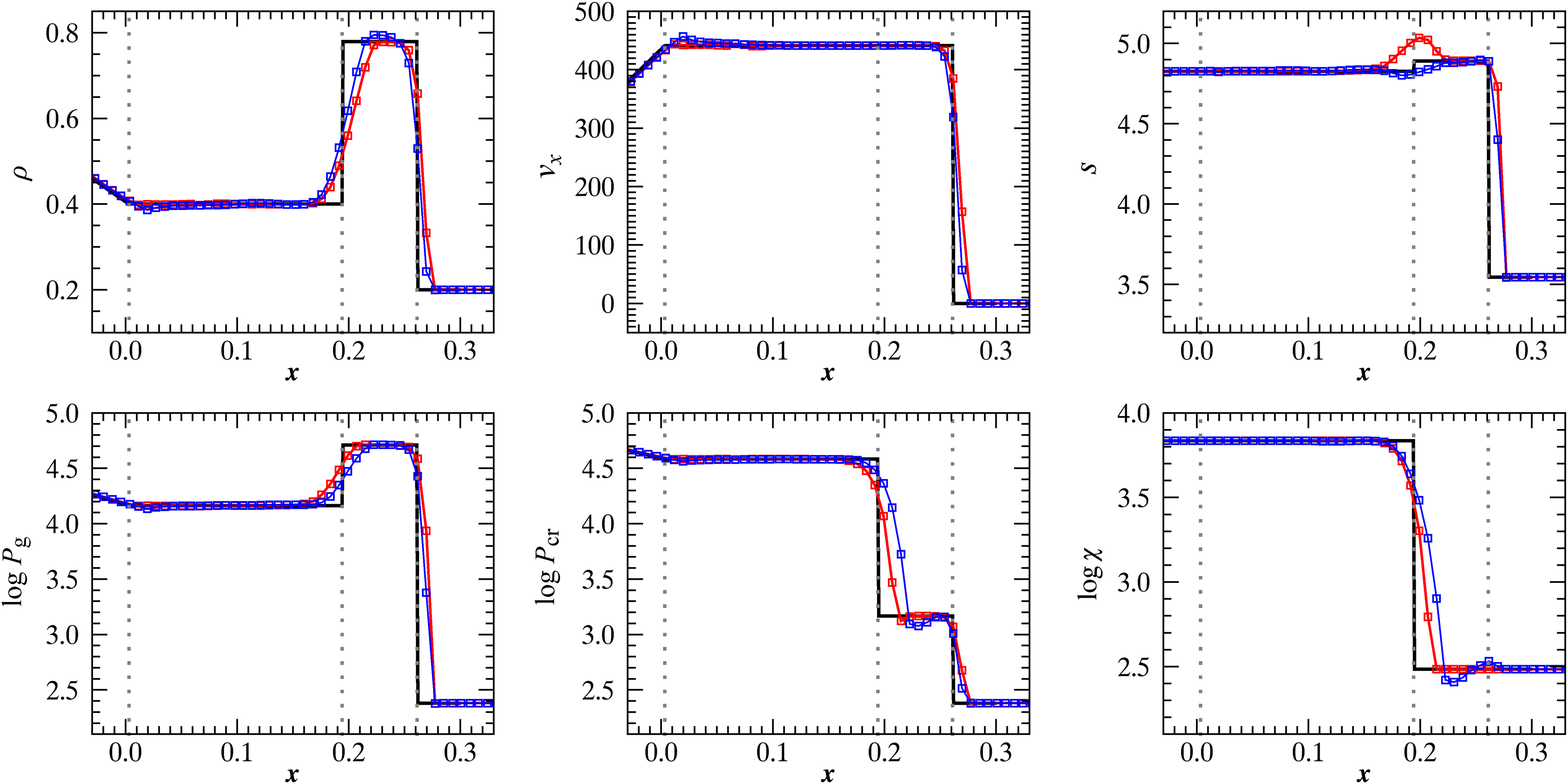}
\end{center}
\caption{\label{P06}  
The CRHD shock tube problem of \protect \cite{2006MNRAS.367..113P}.
The grey dotted lines denote the tail of rarefaction ($x= 3.232 \times 10^{-3}$), contact discontinuity 
($x= 0.1942$), and the shock ($x= 0.2612$) at $  t = 4.4 \times 10^{-4}$ from left to right. 
The CFL number is taken to be 0.794.
}
\end{figure*}    %%%%%%%%%%%%%%%%%%%%%%%%%%

%%%%%%%%%%%%%%%%%%%%%%%%%%%%%%%%%%%%%%%%%%%
\section{Shock Tube Problems} 

We reexamine the shock tube problems examined in earlier works. 

Fig.~\ref{D15} shows the solutions of the shock tube problem of 
\cite{2016A&A...585A.138D} of which initial state is summarized 
in Table~\ref{dubois15}.  We have solved the problem on the
uniform cell width of $ \Delta x = 1/128 $ with the time step, $ \Delta t = 2.45 \times 10^{-3} $.
The blue and red curves denote the solutions at $ t = 0.245 $ 
obtained with Pdv scheme and ours, respectively.   The exact solution
is denoted by the black.   The upper panels denote $ \rho $, $ v _x $ and 
$ s $ as a function of $ x $ from left to right, while the lower panels do 
$ P _{\text{g}} $, $ P _{\text{cr}} $, and $ \chi $.  Only the region of
$ -0.05 \le x \le 0.50 $ is shown.   The Pdv scheme gives an apparently
good approximation for $ \rho $, $ v _x $, $ s $, $ P _{\text{g}} $, and
$ P _{\text{cr}} $ but not for $ \chi $.  Our scheme provides a better
approximation.

The spurious sound wave emission is not observed in this test problem since
$ \gamma _{\text{g}} = \gamma _{\text{cr}}  = 1.4 $.   This is because the spurious 
increase in the pressure is proportional to $ \gamma _{\text{g}} - \gamma _{\text{cr}} $ 
as shown in equation (\ref{PTdif}).

Fig.~\ref{P06} shows the solutions of the shock tube problem 
\cite{2006MNRAS.367..113P} of which initial state is given in Table~\ref{pfrommer06}.
Also this problem has been solved on the uniform cell width of $ \Delta x = 1/128 $ with
the time step, $ \Delta t =8.0 \times 10^{-6}  $.   The solution at $ t = 4.4 \times 10^{-4}$ 
obtained with the Pdv scheme is  denoted by the blue curves, while that obtained 
with our scheme is by the red ones.  The black curves denote the exact solution.
Each panel shows $ \rho $, $ v _x $, $ s $, $ \log P _{\text{g}} $, $ \log P _{\text{cr}} $,
and $\log \chi $.  In this example, both the schemes provide a good approximation.
This is mainly because the CR pressure is by a factor of 10 lower than the gas pressure 
between the contact discontinuity and the shock front ($ 0.1942 \le x \le 0.2612 $).

% Don't change these lines
\bsp	% typesetting comment
\label{lastpage}
\end{document}